\documentclass[12pt,letterpaper]{article}
\usepackage{CJKutf8}
\pdfoutput=1
\usepackage{booktabs}
\usepackage{siunitx}
\usepackage{tikz}
\usepackage{geometry}
\usepackage{tikz}
\usepackage{color}
\usepackage{tcolorbox}
\usepackage{slashed}
\definecolor{refkey}{rgb}{1,0,0}
\definecolor{labelkey}{rgb}{0,0,1}
\usepackage{bbold,amsfonts,amssymb,amsmath}
\usepackage{esint}
\usepackage{rotating}
\usepackage[utf8]{inputenc}
\usepackage{cite}
\usepackage{graphicx}
\usepackage{caption}
\usepackage{subcaption}
\usepackage{booktabs}

\usepackage{listings}

\usepackage{hyperref}

\usepackage{soul}
\setstcolor{red}

\newcommand{\be}{\begin{equation}}
	\newcommand{\ee}{\end{equation}}
\newcommand{\ben}{\begin{displaymath}}
	\newcommand{\een}{\end{displaymath}}
\newcommand{\bea}{\begin{equation}\begin{aligned}}
		\newcommand{\eea}{\end{aligned}\end{equation}}
\newcommand{\bean}{\begin{eqnarray*}}
	\newcommand{\eean}{\end{eqnarray*}}

\newcommand{\bra}[1]{\mbox{$\langle #1 |$}}
\newcommand{\ket}[1]{\mbox{$| #1 \rangle$}}

\newcommand{\eg}{{\it e.g.}}
\newcommand{\ie}{{\it i.e.}}

\newcommand{\tr}{\mbox{Tr}}

\newcommand{\commentout}[1]{}

\renewcommand{\theequation}{\arabic{section}.\arabic{equation}}

\newcommand{\beq}{\begin{equation}}
	\newcommand{\eeq}{\end{equation}}
\newcommand{\beqr}{\begin{displaymath}}
	\newcommand{\eeqr}{\end{displaymath}}
\newcommand{\beqa}{\begin{eqnarray}}
	\newcommand{\eeqa}{\end{eqnarray}}
\newcommand{\beqar}{\begin{eqnarray*}}
	\newcommand{\eeqar}{\end{eqnarray*}}
\newcommand{\cM}{{\cal M}}

\newcommand{\cO}{{\cal O}}
\newcommand{\cF}{{\cal F}}
\newcommand{\cL}{{\cal L}}

\newcommand{\non}{\nonumber}

\newcommand{\half}{\ensuremath{\frac{1}{2}}}
\newcommand{\mpi}{\ensuremath{m_{\pi}}}
\newcommand{\fpi}{\ensuremath{f_{\pi}}}

\renewcommand{\Re}{\ensuremath{\mathrm{Re}}}
\renewcommand{\Im}{\ensuremath{\mathrm{Im}}}

\newcommand{\MeV}{\ensuremath{\mbox{MeV}}}
\newcommand{\GeV}{\ensuremath{\mbox{GeV}}}

\newcommand{\chiPT}{\ensuremath{\chi_{PT}}}

\makeatletter
\newcommand{\hathat}[1]{%
\begingroup%
\let\macc@kerna\z@%
\let\macc@kernb\z@%
\let\macc@nucleus\@empty%
\hat{\raisebox{.2ex}{\vphantom{\ensuremath{#1}}}\smash{\hat{#1}}}%
\endgroup%
}
\makeatother

\definecolor{gblue}{RGB}{15,10,220}
\definecolor{gred}{RGB}{220,10,15}
\newcommand{\gtbblue}[1]{\color{gblue}#1}

\begin{document}

\title{\Large \bf The Gauge Theory Bootstrap: \\ 
Predicting pion dynamics from QCD}

\author{
	Yifei He$^\text{1}$,
	Martin Kruczenski$^\text{2}$ \thanks{E-mail: \texttt{yifei.he@ens.fr, markru@purdue.edu.}} \\
[2.0mm]
$^1$ \small Laboratoire de Physique de l'\'Ecole Normale Supérieure, ENS, Université PSL,\\
\small CNRS, Sorbonne Université, Université Paris Cité, F-75005 Paris, France \\
$^2$ \small Department of Physics and Astronomy and PQSEI\thanks{Purdue Quantum Science and Engineering Institute}, \\
\small Purdue University, West Lafayette, IN 47907, USA.}

\date{\today}

\maketitle

\begin{abstract}

The Gauge Theory Bootstrap \cite{PhysRevLett.133.191601,PhysRevD.110.096001,He:2024nwd} computes the strongly coupled pion dynamics by considering the most general scattering matrix, form factors and spectral densities and matching them with perturbative QCD at high energy and with weakly coupled pions at low energy. In this work, we show that further constraints on the spectral densities significantly reduce the possible solutions to a small set of qualitatively similar ones. Quantitatively, the precise solution is controlled by the asymptotic value of the form factors and SVZ sum rules. We also introduce an iterative procedure that, starting from a generic feasible point, converges to a unique solution parameterized by the UV input. For the converged solution we compute masses and widths of resonances that appear, scattering lengths and effective ranges of partial waves, low energy coefficients in the effective action. Additionally, we use these results to discuss the thermodynamics of a pion gas including pair correlations of pions with same and opposite charge.
\end{abstract}

\clearpage

\tableofcontents

\newpage
\section{Motivation and summary}

In asymptotically free gauge theories, the UV is understood in perturbation theory, and the IR is understood as a weakly coupled theory of pions determined by symmetries and the mass $\mpi$ and coupling $\fpi$ of the pion. In the intermediate energy region both descriptions are strongly coupled and the dynamics is difficult to solve. That region has a rich dynamics known from phenomenological analysis of experiments \cite{Colangelo:2001df,Pelaez:2004vs,Bijnens:1994ie,Bijnens:1997vq,Bijnens:1998fm,Ananthanarayan_2004,Guerrieri:2024jkn,Pelaez:2024uav} and from lattice simulations \cite{FlavourLatticeAveragingGroupFLAG:2024oxs}.

In \cite{PhysRevLett.133.191601,PhysRevD.110.096001,He:2024nwd} we proposed the Gauge Theory Bootstrap (GTB), a method to understand this region from a purely theoretical point of view. The idea is to write the most general S-matrix, form factors and current spectral densities that satisfy basic physical properties and add more and more constraints based on matching of the physics to the known UV and IR. In that way we expect to converge to a unique solution parameterized by the UV and IR properties.  In this paper we get much closer to this main goal by adding further IR constraints on the spectral density that greatly reduce the space of solutions and, further, by creating an iterative procedure that pushes the constraints towards saturation and converges to a unique solution. Instead of using a functional to map out the space of solutions, and extract extremal amplitudes as we did in previous work, we can remove the functional and allow the program to choose an arbitrary feasible point and then use this iterative procedure to converge to a solution. This differs from the standard S-matrix bootstrap procedure that uses kinematic constraints to put bounds on theory parameters, instead we introduce {\it dynamical information}\footnote{We are grateful to A. Vainshtein for emphasizing this point.} from the UV and IR to compute the S-matrix. In that way, we perhaps go back to the original bootstrap philosophy of the 1960's \cite{Eden:1966dnq,chew1966analytic} but augmented with the microscopic dynamical information that was not known at that time. In that work \cite{Eden:1966dnq,chew1966analytic} the dynamical information was the idea of self-consistency, or hadron democracy: for example the $\rho$ meson gave rise to an attractive potential between pions which in turn gave rise to a resonance, the $\rho$ meson itself (see for example \cite{PhysRevLett.7.112,PhysRev.142.1163}). Although interesting, such ideas did not match the experimental results and were superseded by QCD. In the recent bootstrap revival \cite{Paulos:2016fap,Paulos:2016but,Paulos:2017fhb,Kruczenski:2022lot}, powerful numerical methods were used to put bounds on couplings and other parameters of generic quantum field theories, including the effective field theory of pions \cite{Guerrieri:2018uew,Bose:2020cod,Bose:2020shm,Guerrieri:2020bto,Albert:2022oes,Fernandez:2022kzi,Albert:2023seb,Albert:2023jtd,Ma:2023vgc,Bhat:2023puy}. This also extended to form factors and spectral densities (mostly in 2d theories) \cite{Karateev:2019ymz,Karateev:2020axc,Chen:2021pgx,Correia:2022dyp,Cordova:2023wjp}. 
However it is clear that general kinematical constraints, albeit non-trivial, still allow an enormous number of theories and that dynamical information, here provided by QCD, is necessary to identify the correct theory.\footnote{In \cite{Chen:2021pgx}, dynamical information about the $\phi^4$ theory was extracted using Hamiltonian truncation and used as input in bootstrap; in \cite{Correia:2022dyp}, the central charge of $2d$ Ising CFT is used as an input.} In that way, without using experimental input, we can predict and {\it compute} low energy constants, partial waves, form factors, resonance spectrum, etc. For example in the recent work  \cite{Zahed:2024haa} this method was used as a tool to study the $g-2$ factor contribution from hadrons. Notice also that the only assumption on the spectrum is that there are no bound states. The number of resonances together with their masses and widths are results of the computation based on the UV input from pQCD and matching with low energy pions. 

One caveat that became apparent due to this improved procedure is that, at the moment, the results depends on having accurate values of the form factors at high (but not asymptotically high) energy ($\sqrt{s}\sim 2\,\GeV$). At that energy calculations of the Brodsky-Lepage \cite{osti_1447331} type give the form factors up to a factor of order one. Therefore we allow for a range of values starting from the minimum that the constraints allow. We discuss this later in the paper. On the plus side, the asymptotic form factors is the only place where $\fpi$ is used and as we just mentioned, we allow for a range of values meaning that the precise value of $\fpi$ is not really used. Therefore, $\fpi$ when interpreted as a low-energy coupling is an outcome of our computation.

\smallskip

As we did in our previous work \cite{He:2024nwd}\footnote{A frozen snapshot of the numerical program for that paper is available at \url{https://github.com/hyfysics/gauge-theory-bootstrap/releases/tag/arxiv-2403.10772}}, we have made publicly available the numerical program used to obtain the results of this paper at \url{https://github.com/hyfysics/gauge-theory-bootstrap/releases/tag/arxiv-2505.19332}. The latest version of the numerical GTB program can be found in the GitHub repository: \href{https://github.com/hyfysics/gauge-theory-bootstrap}{\color{blue}gauge-theory-bootstrap}.

\section{New ingredients to the GTB procedure}

The procedure we use in this paper is basically the same as described in previous work and we refer the interested reader to \cite{PhysRevLett.133.191601,PhysRevD.110.096001,He:2024nwd} for necessary details and main notation\footnote{See \cite{Cordoba:2025wnv} for an alternative numerical implementation.}. In this section we discuss two crucial new additions to the GTB procedure. First we introduce, from $\chiPT$, constraints on the low energy kinematic (\ie\ from the free pion Lagrangian) behavior of the spectral densities using a parameterization that we discuss in the appendix. Second, and perhaps the most importantly, we devise an iterative procedure that starts from an arbitrary feasible point and converges to a unique solution (after fixing the UV and IR information). We now discuss those new ingredients in detail as well as the input from the asymptotic form factors. Once more, the only numerical input to the GTB procedure are the values of $\mpi$ that sets the units, $\fpi$ that is used in the asymptotic form factors, $\alpha_s$, $N_c=3$, $N_f=2$ and $m_q$ from the gauge theory to compute the SVZ sum rules \cite{SHIFMAN1979385,SHIFMAN1979448,Shifman:1978bw}. At the moment, however, all pQCD calculations are done with $m_q=0$ assuming $m_q\ll \Lambda_{QCD}$. Therefore $m_q$ only appears as an overall factor in the S0 sum rule. Presumably $m_q$ should be computed by consistency in future work. The fact that  $\fpi$ does not appear in the low energy matching, and only in the asymptotic form factors is discussed around  eq.\eqref{chiralconstraints}.

\subsection{Low energy spectral density}\label{rholowE}

A major role in the Gauge Theory Bootstrap is played by the currents\footnote{It is customary to call QCD operators currents whether they are actual currents or not. Of course these operators play an important role in most studies of QCD, not just here.}, a set of QCD operators chosen to match the quantum numbers of the different partial waves. In this and previous papers we consider the following operators for $N_f=2$ 
\begin{subequations}
\beqa
j_S &=& m_q(\bar{u}u+\bar{d}d) = \frac{1}{2}m^2_{\pi}\pi^a\pi^a +\cO(\pi^4) \label{jscalar}\\
j_V &=& \frac{1}{2}(\bar{u}\gamma^\mu u-\bar{d}\gamma^\mu d)\Delta_\mu  = \epsilon^{abc} \pi^b \partial_\Delta \pi^c +\cO(\pi^4) \\
j_T &=& T^{\mu\nu} \Delta_\mu\Delta_\nu  
\eeqa
\end{subequations}
written both from QCD and from the free pion Lagrangian. Here $T^{\mu\nu}$ is the energy momentum tensor (either in QCD or in terms of pions) and $\Delta_\mu=(0,1,i,0)$ is a (complex) light-like vector $\Delta^2=0$ with components in the $(x,y)$ plane. 
In general, currents,  can be identified at low energy up to a normalization constant, however, for these three, the normalization is known. Using the free pion Lagrangian, namely kinematically, one can compute the spectral densities at low energy \cite{GASSER198765}:
\begin{subequations}
\beqa
\rho^0_0(s) &\simeq& \frac{m_{\pi}^4}{(2\pi)^4}\, \frac{3}{16\pi}\,\left(1-\frac{4m_{\pi}^2}{s}\right)^{\frac{1}{2}} \\ 
\rho^1_1(s) &\simeq& \frac{1}{(2\pi)^4}\,\frac{s}{24\pi}\, \left(1-\frac{4m_{\pi}^2}{s}\right)^{\frac{3}{2}} \label{vecrho}\\ 
\rho^0_2(s) &\simeq&  \frac{1}{(2\pi)^4}\,\frac{s^2}{160\pi}\left(1-\frac{4m_{\pi}^2}{s}\right)^{\frac{5}{2}}
\eeqa
\end{subequations}
In appendix \ref{rhopara} we describe a parameterization of the spectral densities that takes into account this behavior.  This is the first addition to the procedure and plays a crucial role because the spectral density is now also constrained in the IR and not only in the UV (by the SVZ sum rules as explained in section \ref{svz1} and appendix \ref{fesr3}). 

\subsection{Iterative ``Watsonian" procedure}\label{sec:it}

\begin{figure}[t]
 	\centering
 	\includegraphics[width=\textwidth]{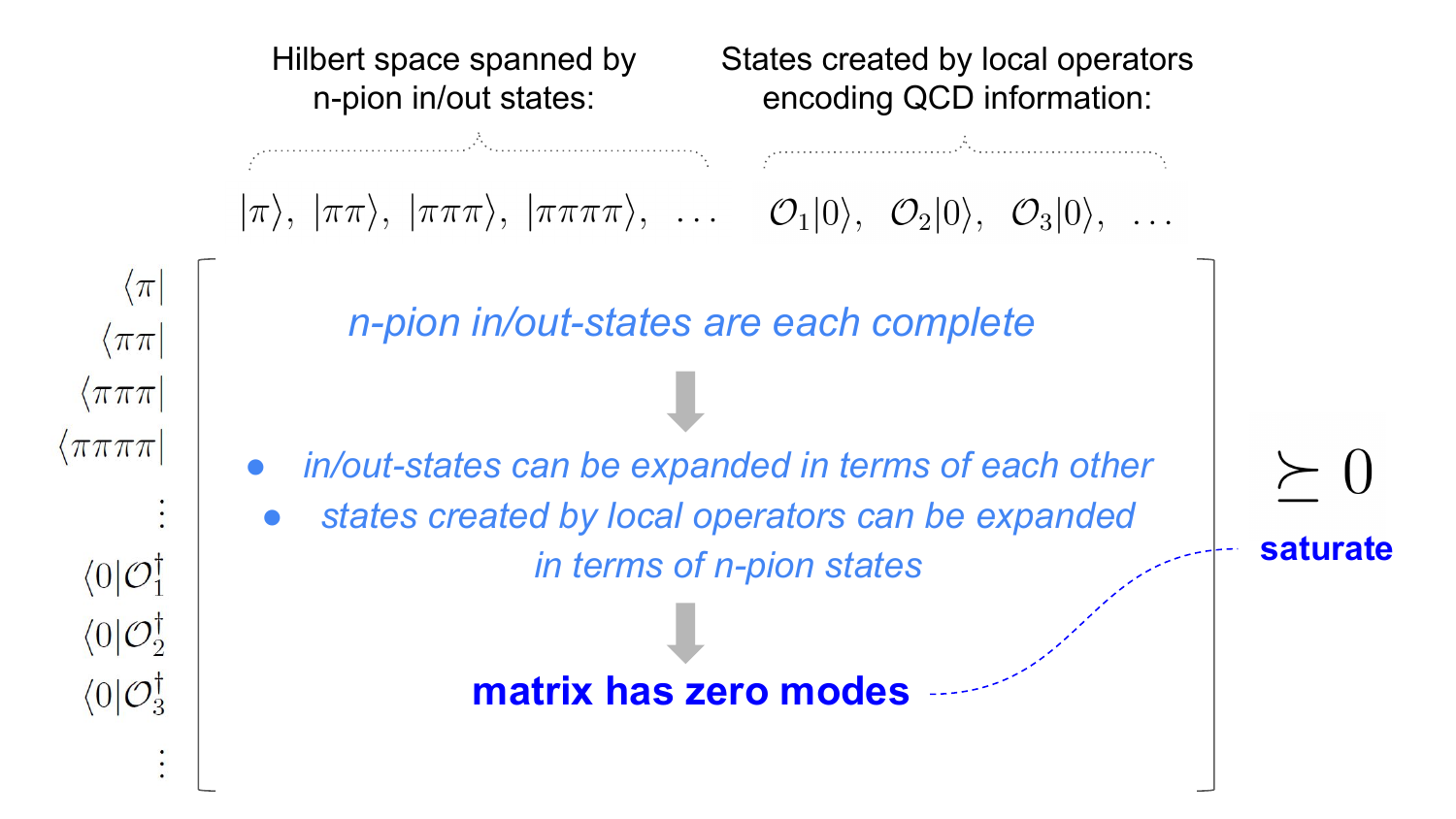}
 	\caption{The matrix of state overlaps should saturate the positive semidefinite condition and have zero modes.}
 	\label{zeromode}
 \end{figure}

An important ingredient of any bootstrap program are positivity constraints associated with the positivity of the norm in Quantum Mechanics, for example in the Gauge Theory Bootstrap we use \eqref{Bmt} below. If a large enough matrix, with many pion states and QCD operators included is considered, we expect such matrix of state overlaps to saturate the positivity condition since scattering is unitary and QCD operators acting on the vacuum should be representable as linear combinations of pion states giving rise to zero modes or null states. See fig. \ref{zeromode} for an illustration.
 In \eqref{Bmt} below, where only two pion states are considered, this saturation should certainly be exact below the four pion threshold and approximate above. In general, we adopt the philosphy of always look for saturation of the constraints (i.e. matrix with zero modes) and, if that is not possible to further add pion states and QCD operators to the matrix. Once a matrix such as  \eqref{Bmt} is chosen, to achieve the desired saturation, we now introduce an iterative procedure that leads to a particular solution to the GTB constraints.

Iterative procedures in the S-matrix bootstrap were introduced initially in \cite{He:2018uxa} to get closer to the boundary of the allowed space. The main idea is that, denoting the bootstrap variables as $x_j$, and given a constraint $\Phi_A(x_j) \le 0$, after finding a solution $x_j=\bar{x}_j$, we can improve the solution to saturate the constraint by maximizing again in the direction of the gradient of the constraint, namely a functional linear in the variables $x_j$:
\begin{equation}
\mathcal{F}=\sum_{A,j}\left.\frac{\partial \Phi_A}{\partial x_j}\right|_{\bar{x}_j}x_j
\end{equation}
In terms of the normalized partial wave $h^I_\ell(s)$ where 
\beq\label{hdef}
S^I_\ell(s) = 1+ih^I_{\ell}(s)=\eta^I_{\ell}(s) e^{2i\delta_\ell^I(s)}
\eeq
with $\delta_\ell^I$ the usual phase shift, the unitarity constraint reads
\beq\label{huni}
(\Re h^I_\ell(s))^2 + (\Im h^I_\ell(s))^2 - 2 \Im h^I_\ell(s) \le 0,\;\; s>4m^2_{\pi},\;\forall\;I,\ell
\eeq
It is known, qualitatively, that 2-to-2 pion scattering tends to saturate  unitarity even at energies larger than the four pion threshold.\footnote{at higher energy we could incorporate some method of particle production such as was done in \cite{Tourkine:2021fqh,Tourkine:2023xtu} or via resonance decay as suggested in \cite{He:2024nwd}} If we want to approach further the unitarity saturation, we can linearize the constraint \eqref{huni} and the new functional to maximize, for the next iteration, is
\beq\label{newfunc}
\cF = \int ds \sum_{\ell,I} \left.\Re h^I_\ell(s)\right|_{old}\, \left.\Re h^I_\ell(s)\right|_{new} + \left.\Im h^I_\ell(s)\right|_{old}\, \left.\Im h^I_\ell(s)\right|_{new} -  \left.\Im h^I_\ell(s)\right|_{new} 
\eeq
where {\it old} are the values from the previous iteration, and {\it new} are the values in the maximization that is being performed. In the case of two dimensions it was very successful in converging to the desired solutions. One might expect that the same happens here. And although this is true, once we include form factors and spectral densities in the bootstrap, Watson theorem provides a {\it new natural iterative functional} for our purpose. Watson theorem establishes that, when four-pion state contribution is negligible, the 2-pion form factor can be written as
\begin{equation}
    \mathcal{F}^{I}_{\ell}={}_{\text{out}}\langle\pi\pi|\mathcal{O}^I_{\ell}|0\rangle={}_{\text{out}}\langle\pi\pi|\pi\pi\rangle_{in}\;{}_{in}\langle\pi\pi|\mathcal{O}^I_{\ell}|0\rangle=S^I_{\ell}\mathcal{F}^{I*}_{\ell}
\end{equation}
so the phase of the form factors agrees with the phase shift of the corresponding partial wave. The procedure then replaces the phase of the previous (old) partial wave by the phase of the previous (old) form factor when computing the functional, {\it i.e.}, in \eqref{newfunc} we replace:
\beq
\left. h^I_\ell(s)\right|_{old} \rightarrow \left.\left(|h^I_{\ell}(s)| \frac{F^I_\ell(s)} {|F^I_\ell(s)|}\right)\right|_{old}
\eeq
Namely instead of just using the old partial wave to determine the functional, we use its modulus and replace its phase by the phase of the form factor. Such iteration will tend to converge to unitarity saturation while matching the phase of the form factor with the phase shift. From a physical point of view, the UV information incorporated into the form factors and spectral densities is conveyed to the phase shifts and partial waves. At the same time the crossing and unitarity constraints on the partial waves feed back to the form factors. 

This iteratative procedure can be directly interpreted from the zero modes of the matrix. Following the form factor bootstrap
\cite{Karateev:2019ymz}, we introduce a $3\times 3$ positive semidefinite matrix of inner products:
   \beq\label{Bmt}
  G_{\text{SDP}}= \begin{array}{c c} &
   	\begin{array}{c c c} \ket{\mbox{in}}_{P,I,\ell} & \ket{\mbox{out}}_{P,I,\ell} & \cO_{P,I,\ell}\ket{0} \\
   	\end{array}
   	\\
   	\begin{array}{c c c}
   		\bra{\mbox{in}}_{P',I,\ell}  \\
   		\bra{\mbox{out}}_{P',I,\ell}\\
   		\bra{0} \cO^\dagger_{P',I,\ell}
   	\end{array}
   	&
   	\left(
   	\begin{array}{c c c}
   		1 &\ S^I_\ell(s)\ &\ \ \cF^I_\ell(s) \\
   		S_\ell^{I*}(s) &\ 1\ & \ \ \cF^{I*}_\ell(s) \\
   		\cF^{I*}_\ell(s) &\ \cF^I_\ell(s)\ &\ \ \rho^I_\ell(s)
   	\end{array}
   	\right) 
   \end{array} \succeq 0,\;\; \forall I,\ell, \ s>4m^2_{\pi},\;\; 
   \eeq
   where we removed an overall factor $\delta^4(P-P')$, namely the identity in the center of mass momentum. The in- and out- states are two pion states and $\cO_{P,I,\ell}$ is a momentum space current. The elements of the matrix are the scattering matrix $S_\ell^I$, the form factors $\cF_\ell^I$ and the spectral density $\rho_\ell^I$. Recall that for positive semidefinite Hermitian matrix $G_{\text{SDP}}$, one has
\begin{equation}\label{funal}
   \text{Tr}(M\cdot G_{\text{SDP}})\ge 0,\;\; \forall\; M=vv^{\dagger}
\end{equation}
where $v$ is an arbitrary complex vector.
When the positive semidefinite condition is saturated, this matrix has two zero eigenvalues. That means that two linear combinations of the states vanish. One simply means that the in and out states are the same (up to a phase) if there is no particle production. The other, interestingly, means that we have a way to write $\cO_{I,\ell,P}\ket{0}$ as a two pion state.    So, from a physical perspective, one can understand the zero modes as a way to write pQCD currents acting on vacuum in terms of pion states and vice versa. In terms of the matrix elements, this gives rise to the conditions:
\begin{equation}
    S^I_{\ell}=\frac{\mathcal{F}^I_{\ell}}{\mathcal{F}^{I*}_{\ell}},\;\; \rho^I_{\ell}=\mathcal{F}^I_{\ell}\mathcal{F}^{I*}_{\ell}
\end{equation}
where the first one is equivalent to the Watson theorem.
The matrix $\eqref{Bmt}$ then becomes
   \begin{equation}\label{Gnull}
  G_{\text{sat}}= \begin{pmatrix}
        1&\frac{\mathcal{F}_{\ell}(s)}{\mathcal{F}^*_{\ell}(s)}&\mathcal{F}_{\ell}(s)\\
\frac{\mathcal{F}^*_{\ell}(s)}{\mathcal{F}_{\ell}(s)}&1&{\mathcal{F}^*_{\ell}(s)}\\
{\mathcal{F}^*_{\ell}(s)}&{\mathcal{F}_{\ell}(s)}&{\mathcal{F}_{\ell}(s)}{\mathcal{F}^*_{\ell}(s)}
    \end{pmatrix}
\end{equation}
with two zero modes
\begin{equation}\label{v12}
v_1=\begin{pmatrix}-\mathcal{F}_\ell(s)\\
    0\\
    1
    \end{pmatrix},\;\; v_2=\begin{pmatrix}-\frac{\mathcal{F}_\ell(s)}{\mathcal{F}^*_\ell(s)}\\
    1\\
    0
    \end{pmatrix}
\end{equation}
and we have
\begin{equation}
    \text{Tr}(M_i\cdot G_{\text{sat}})=0,\;\; M_i=v_iv_i^{\dagger},\;\; i=1,2
\end{equation}
i.e., the minimum attainable for \eqref{funal}.
Now, given one old solution $\mathcal{F}^{(\text{old})}$, we can construct the matrices $M^{(\text{old})}=v_iv_i^{\dagger}$ following \eqref{v12}, and write down a linear functional $\cF_P$ to minimize
\begin{equation}
\text{min}\; \left\{ \cF_P =\int ds\sum_{I,\ell}\text{Tr}\big(M^{(\text{old})}\cdot G^{(\text{new})}\big) \right\}
\end{equation}
Repeating this procedure until convergence leads to the GTB solution, i.e., the solution we look for is a fix point of this iterative procedure. Taking  $i=2$, the functional becomes precisely the Watsonian iterative functional we describe above.\footnote{In addition, we can also take the other zero mode $i=1$ which we leave for future work.} This procedure should be interpreted as projecting onto the null space of the positive semidefinite matrix iteratively.

\subsection{Asymptotic form factors}\label{chifact}

This new improved procedure highlights the need for more precise high energy inputs. This is not a problem for the SVZ sum rules \cite{SHIFMAN1979385,SHIFMAN1979448,Shifman:1978bw} but it is a problem for the form factors. The theoretical computation of form factors at high energy is well established \cite{MUELLER1981237} and proceeds similarly to the better known case of Deep Inelastic Scattering. However, there is less validation from experiments specially in the case of the pion.   
Qualitatively, at high energy, the pion is highly boosted but in the transverse directions it has the same size given by its radius $R_\pi \sim \frac{1}{\fpi}$. The probability of a probing particle hitting a quark is inversely proportional to the transverse area. That is the reason that $\fpi$ appears in the asymptotic formulas. More technically, the distribution amplitude is required, a low energy information. At high energy the results of Brodsky-Lepage \cite{osti_1447331,Tong_2021, Tong_2022,He:2024nwd} allows us to determine form factors as
\begin{subequations}
\beqa
|F_1^1(s)| & \simeq & \frac{16\pi\alpha_s f_\pi^2}{s}\\
|F_2^0(s)| & \simeq & \frac{48\pi\alpha_s f_\pi^2}{s}
\eeqa
\end{subequations}
The scalar form factor was crudely estimated in \cite{He:2024nwd} to be $|F_0^0(s)|  \sim   \frac{1}{s} \ln\left(\frac{s}{\mpi^2}\right)$. At an energy $s_0 \sim 2\GeV$ these formulas are not quite valid. For example it has been argued that the $P1$ form factor changes by a factor of up to 3 \cite{PhysRevLett.111.141802}. So we define $\chi_{\ell}^I$, three rescaling factors of order one and require:  
\begin{subequations}
\beqa
|F_0^0(s>s_0)| & \le & \chi_0^0\ \frac{\mpi^4}{s}\, \ln\!\left(\frac{s}{\mpi^2}\right) \\
|F_1^1(s>s_0)| & \le  & \chi_1^1\ \frac{16\pi\alpha_s f_\pi^2}{s}\\
|F_2^0(s>s_0)| & \le  & \chi_2^0\ \frac{48\pi\alpha_s f_\pi^2}{s}
\eeqa
\end{subequations}
where in the scalar case we just chose the factor $\mpi^4$ from units.
If we choose any of these factors $\chi^I_{\ell}$ to be one, the problem becomes unfeasible reflecting the fact that at $s_0$ we are not in the fully asymptotic region yet. Most plots in this paper are done with the values \beq\label{chichoice}
\chi_0^0=8,\; \; \chi_1^1=2.2,\; \; \chi^0_2=9,
\eeq
which are close to the minimum allowed values. The partial waves and form factors seems to agree reasonable well with the experimental results but this should not be taken as a strong prediction of the procedure. Later in section \ref{sec:FF} we show how the solution changes if we change these values over a range (see figs. \ref{rhomass1}, \ref{psffchi}, \ref{rhomass2}, \ref{rhomass3}). We emphasize that we are not just computing masses and widths of resonances but the whole amplitude and form factors as analytic functions. We do not assume how many resonances appear in each channel or any other qualitative information, so the overall picture is a result of the calculation. Also, notice again that this is the only place where the value of $\fpi$ enters. 

\subsection{SVZ sum rules}\label{svz1}

As proposed in \cite{PhysRevLett.133.191601,PhysRevD.110.096001,He:2024nwd}, important information about the gauge theory is incorporated into the bootstrap by means of the SVZ sum rules \cite{SHIFMAN1979385,SHIFMAN1979448,Shifman:1978bw}. In this paper, as compared to the previous ones we use a slightly improved version of the finite energy sum rules as is explained in detail in Appendix \ref{fesr3} where we also quote 3-loops results taken from the available literature on perturbative QCD. 

\section{Convergence of the solution}
To test this new improved procedure, we fix the asymptotic (high energy) form factors and sum rules and then maximize and minimize a given physical quantity. The range of values should be relatively small and the partial waves associated with the maximum and minimum should be qualitatively similar. Furthermore, if we apply the iterative procedure to those initial solutions, or to any other arbitrary solution, the procedure should converge to a certain result. The values of the physical quantities for the given UV/IR input will be estimated by the values after convergence.    

\subsection{Test by computing the pion coupling $\lambda$}

It is customary to define the coupling $\lambda$ as the $\pi^0$ scattering amplitude at the symmetric point:
\beq\label{lambdadef1}
\lambda = \frac{1}{32\pi} \left.\cM(\pi^0\pi^0\rightarrow \pi^0\pi^0) \right|_{s=t=u=\frac{4}{3}m^2_\pi} = \frac{3\pi}{4} A\left(\frac{4}{3}m^2_\pi,\frac{4}{3}m^2_\pi,\frac{4}{3}m^2_\pi\right)
\eeq  
which is bounded from above and below from analycitity, crossing and unitarity: $-8.02\le\lambda\le 2.661$ \cite{LM,Paulos:2017fhb,He:2021eqn,Guerrieri:2021tak,Chen:2022nym,EliasMiro:2022xaa}. For low energy pions it is approximately (tree level amplitude)
\beq\label{lambdaW}
\lambda \simeq \frac{\mpi^2}{32\pi\fpi^2} \simeq 0.023
\eeq
which, as we mentioned in \cite{PhysRevD.110.096001}, is a simple way to argue that the theory of pions is weakly coupled at low energy. 
\begin{figure}[t]
 	\centering
 	\includegraphics[width=0.8\textwidth]{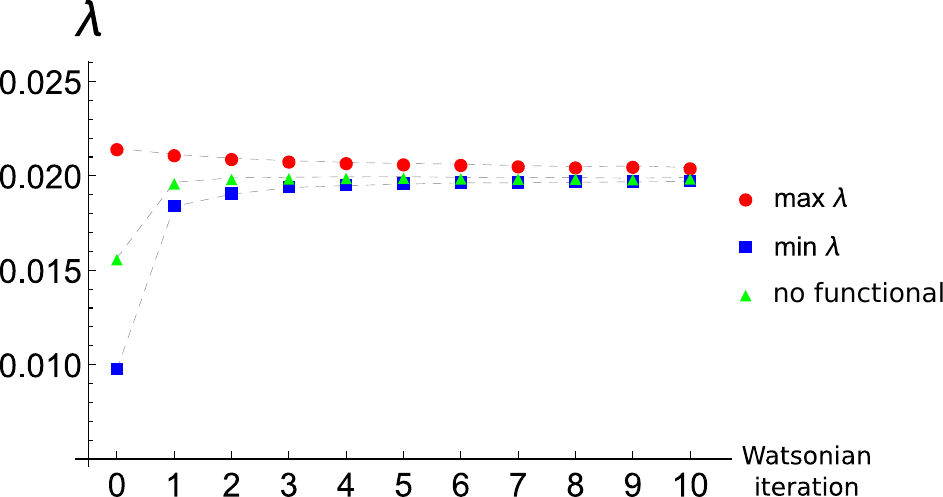}
 	\caption{Convergence of max-min $\lambda$ with Watsonian iterations. maximizing: \protect\tikz[baseline=-0.65ex] \protect\fill[red] (0,0) circle (0.12cm);,
 minimizing: \protect\tikz[baseline=-2.5 ex] \protect\fill[blue] (-0.12,-0.25) rectangle (0.12,-0.5);,
 no functional: \protect\tikz[baseline=-1.2 ex] \protect\fill[green] (0,0) -- (-0.12,-0.2) -- (0.12,-0.2) -- cycle; }
 	\label{lambdatest}
 \end{figure}
  \begin{figure}[t]
 	\centering
 	\includegraphics[width=\textwidth]{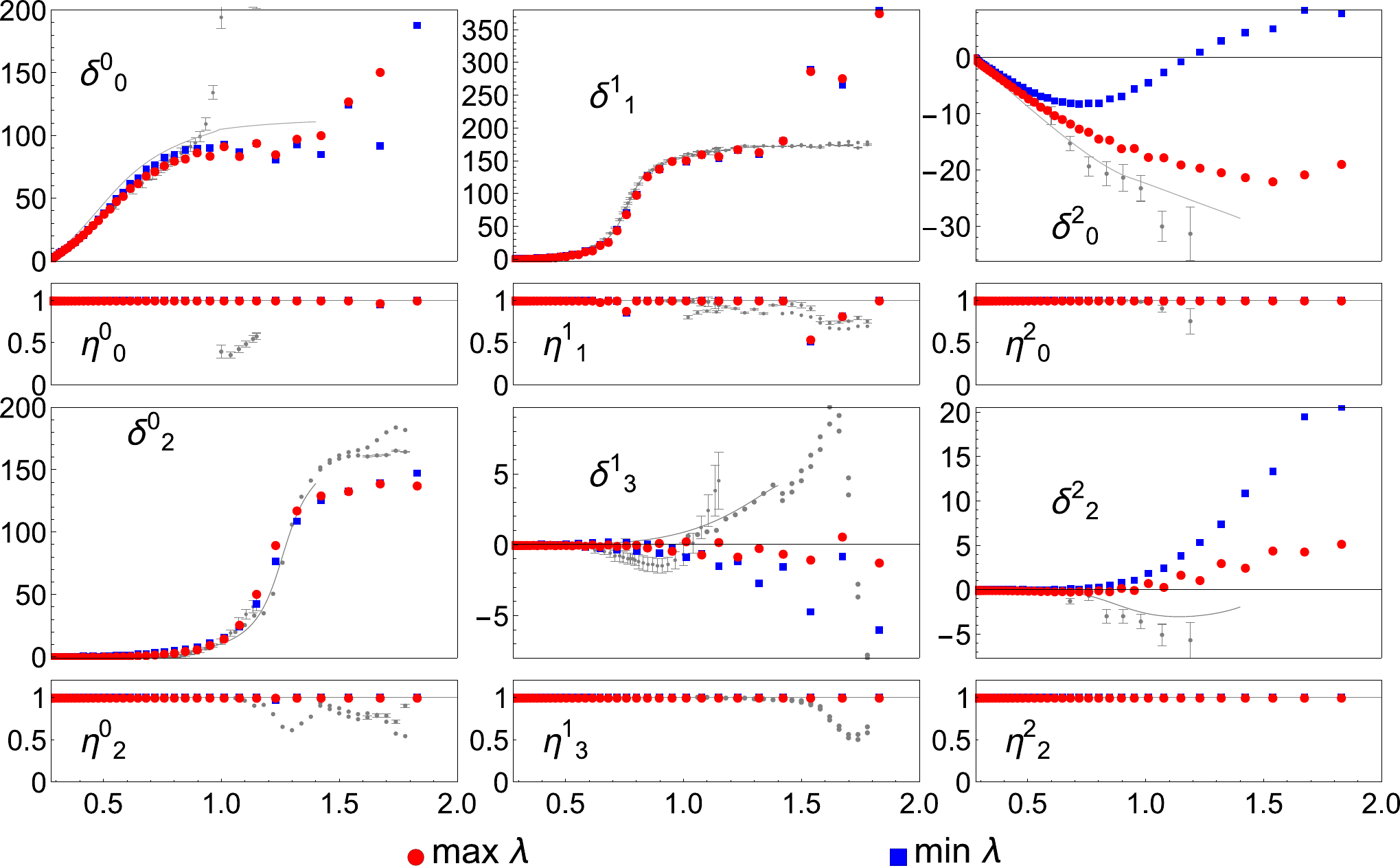}
 	\caption{Partial waves of the maximum and minimum $\lambda$ after 1 Watsonian iteration. As a reference we plot the experimental data from\cite{Protopopescu:1973sh,LOSTY1974185,Hyams:1975mc} (gray dots) as well as phenomenological fits from \cite{Pelaez:2004vs} (gray lines). In the experimental data, the $S0$ wave has a bump from Kaon pair production that we do not have since we do not include the $s$ quark. Resonances are clearly visible in the $S0$, $P1$ and $D0$ channels.}
 	\label{lambdatestpw}
 \end{figure}
In this work the low and high energy constraints severely restrict the values of $\lambda$. Given a set of input parameters,  we maximize and minimize $\lambda$ obtaining an initial spread $0.01\lesssim \lambda \lesssim 0.022$. See fig. \ref{lambdatest}. For the maximum and minimum, we plot the partial waves\footnote{In all cases we always do one initial Watsonian iteration to smoothen out the functions.} in six channels in fig.\ref{lambdatestpw} and see that they are qualitatively similar, except in the isospin $I=2$ channels ($S2$ and $D2$) and the spin 3 ($F1$) channel, a fact that can be understood due to the lack of high energy input in those channels. This tells us that the space of solutions is indeed tightly controlled by the asymptotics as we can also see by changing the asymptotic values in sec.\ref{chichange}. After that, we take the solutions associated with the maximum and the minimum and apply the iterative procedure of section \ref{sec:it}. In figure \ref{lambdatest} we show the convergence of the values of $\lambda$ to a definite one, whether we start from the maximum $\lambda$, from the minimum or from a feasible point chosen by the optimizer (using no functional\footnote{Technically the minimizer program is required to minimize an $\cF= 0$ functional}). So, the final converged value of $\lambda=0.02$ we take as the computed value of $\lambda$ for these values of the UV/IR input.  
 The main thing we see, is that the existence of resonances in the $S0$, $P1$ and $D0$ channels is a generic result of the procedure that was not input in any way. This is mostly due to the SVZ sum rules combined with the asymptotic behavior of the form factors.

\subsection{Test by computing scattering lengths}

\begin{figure}[h!]
 	\centering
 	\includegraphics[width=0.7\textwidth]{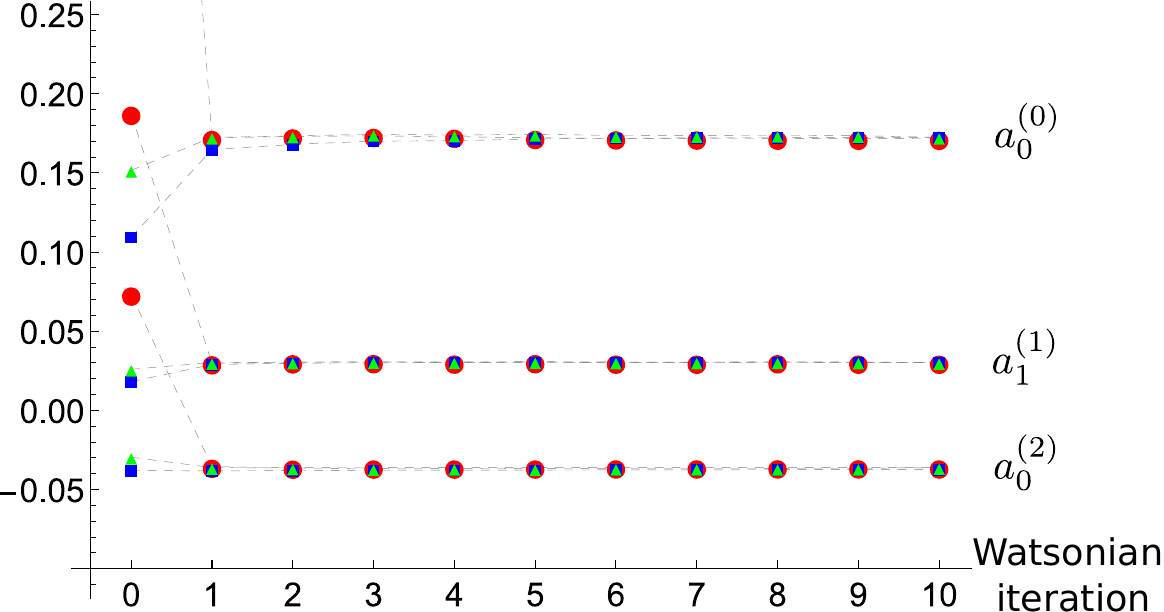}
 	\caption{Convergence of max-min $a^{(I)}_{\ell}$ with Watsonian iterations. maximizing: \protect\tikz[baseline=-0.65ex] \protect\fill[red] (0,0) circle (0.12cm);,
 minimizing: \protect\tikz[baseline=-2.5 ex] \protect\fill[blue] (-0.12,-0.25) rectangle (0.12,-0.5);,
 no functional: \protect\tikz[baseline=-1.2 ex] \protect\fill[green] (0,0) -- (-0.12,-0.2) -- (0.12,-0.2) -- cycle;}
 	\label{atest}
 \end{figure}

Here we repeat the calculation of the previous section for the $S0,P1,S2$ scattering lengths $a^{(0)}_0,a^{(1)}_1,a^{(2)}_0$ defined from:
\begin{equation}
    \text{Re}\, h^I_{\ell}(s)\overset{k\to 0}{\simeq}\frac{4m_\pi}{\sqrt{s}}k^{2\ell+1}\big(a^I_{\ell}+b^I_{\ell}k^2+\ldots\big),\;\; k=\frac{\sqrt{s-4m^2_\pi}}{2}\;.
\end{equation}
where $S^I_\ell = 1+ ih_\ell^I$ as in \eqref{hdef}. 
It should be noted that according to the pure S-matrix bootstrap \cite{Guerrieri:2018uew}, the scattering lengths do not have upper bounds. This seems to be reflected here in the fact that the UV and IR constraints still allow an upper value quite larger than the minimum one. In spite of that, the iteration procedure converges again whether we start from maximum, minimum or just a feasible point. The results are shown in figure \ref{atest}

\subsection{Test by computing pion radii}

\begin{figure}[h!]
 	\centering
 	\includegraphics[width=0.7\textwidth]{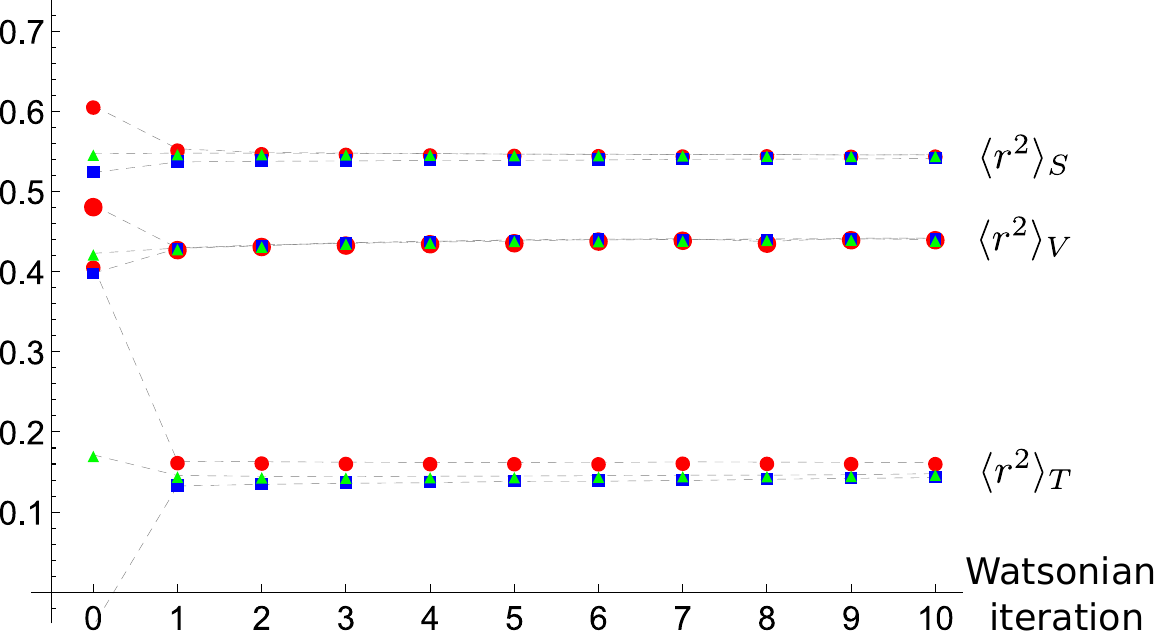}
 	\caption{Convergence of max-min $\langle r_\pi^2\rangle^I_\ell$ with Watsonian iterations. maximizing: \protect\tikz[baseline=-0.65ex] \protect\fill[red] (0,0) circle (0.12cm);,
 minimizing: \protect\tikz[baseline=-2.5 ex] \protect\fill[blue] (-0.12,-0.25) rectangle (0.12,-0.5);,
 no functional: \protect\tikz[baseline=-1.2 ex] \protect\fill[green] (0,0) -- (-0.12,-0.2) -- (0.12,-0.2) -- cycle;. As conventional we denote $\langle r_\pi^2\rangle^0_0=\langle r_\pi^2\rangle_S$,\ $\langle r_\pi^2\rangle^1_1=\langle r_\pi^2\rangle_V$,\ $\langle r_\pi^2\rangle^0_2=\langle r_\pi^2\rangle_T$, the scalar vector and tensor radii.}
 	\label{rtest}
 \end{figure}

In this case, we show in figure \ref{rtest} convergence of the iterations for the pion radii defined from an expansion of the form factor around $s=0$:
\begin{equation}
    F^I_\ell(s)=F^I_\ell(0)\Big(1+\frac{1}{6}\langle r_\pi^2\rangle^I_\ell+\ldots\Big)\label{rdef}
\end{equation}

 In summary, these tests show that the procedure uniquely converges to a solution that satisfies all constraints.

\section{Detailed study of a converged solution}\label{dynamics}

\begin{figure}[h!]
 	\centering
    \includegraphics[width=\linewidth]{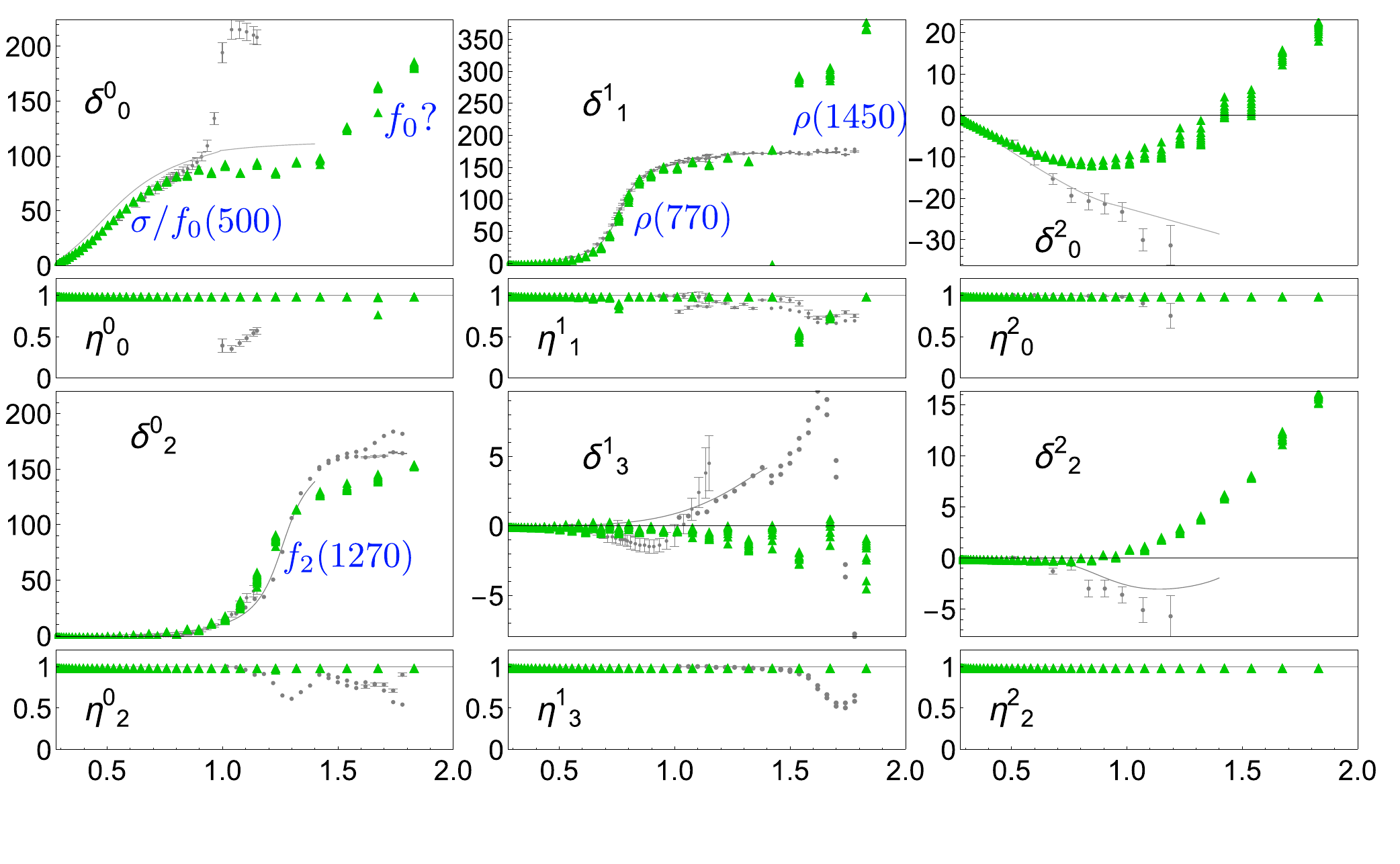}
    \includegraphics[width=\linewidth]{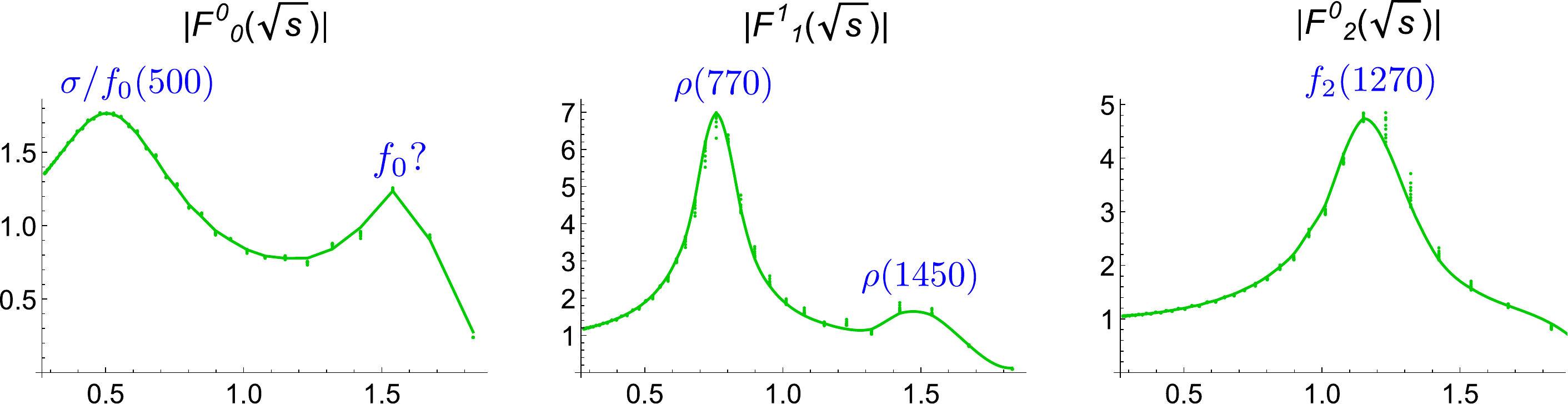}
 	\caption{Partial waves and form factors for 10 iterated solutions (\protect\tikz[baseline=-1.2 ex] \protect\fill[green] (0,0) -- (-0.12,-0.2) -- (0.12,-0.2) -- cycle;) using no-functional, namely starting for an arbitrarily feasible point. Horizontal axes are energy/momentum transfer in $\GeV$. The experimental/phenomenological values in gray are the same as in fig. \ref{lambdatestpw}.}
 	\label{pw}
 \end{figure}

Having shown that, given some fixed UV data, the procedure converges to a unique solution, we
analyze here the properties of the converged solution computing low energy coefficients, scattering lengths, charge radii, etc. In figure \ref{pw}, we show the partial waves and form factors of the converged solution we obtained in the previous section using no functional, as well as comparison with experimental and phenomenological work. In obtaining this, we still keep the values that determine the form factor asymptotics as $\chi_0^0=8$, $\chi_1^1=2.2$, $\chi^0_2=9$ as in \eqref{chichoice}.  In section \ref{chichange}  we will study how the results change with those values.

\subsection{Low energy parameters and one-loop $\chiPT$ amplitude} 

By comparing the pion scattering amplitude $A(s,t,u)$ of the converged solution  with the $\chiPT$ one-loop amplitude (see appendix \ref{chiPT}), we can extract the low energy constants $\bar{\ell}_j$ and the pion coupling $\lambda$, and from there the low energy $\fpi$  using \eqref{fpicalc}. Also, we can match the partial waves in fig. \ref{pw} with the low energy expansion of the partial waves and form factors in \eqref{hdef} and \eqref{rdef} (or use dispersion relations) to extract the scattering lengths, effective ranges and pion radii. The results are presented in tables \ref{tab:LECs1},\ref{tab:LECs2}. 

\begin{table}[h]
    \centering
    \renewcommand{\arraystretch}{1.4}
    \begin{tabular}{c||c
     >{\raggedleft\arraybackslash}p{1.5cm}
        @{$\,\pm\,$}
        S[table-format=1.4]
         >{\raggedleft\arraybackslash}p{1.5cm}
        @{$\,\pm\,$}
        S[table-format=1.6]
         >{\raggedleft\arraybackslash}p{1.2cm}
        @{$\,\pm\,$}
        S[table-format=1.4]}
        \toprule
        & \textbf{GTB} & \multicolumn{2}{c}{\textbf{GL}} & \multicolumn{2}{c}{\textbf{Bij}} & \multicolumn{2}{c}{\textbf{CGL}} \\
        \midrule
        $\bar{\ell}_1$ & \textcolor{blue}{1.6} &  -2.3 &  3.7  & -1.7  &  1.0 &  -0.4  &  0.6  \\
        $\bar{\ell}_2$ & \textcolor{blue}{5.5} &   6.0 &  1.3  &  6.1  &  0.5 &   4.3  &  0.1  \\
        $\bar{\ell}_3$ & \textcolor{blue}{7.8} &   2.9 &  2.4  & \multicolumn{2}{c}{\textbf{}}  &   \multicolumn{2}{c}{\textbf{}} \\
        $\bar{\ell}_4$ & \textcolor{blue}{4.7} &   4.3 &  0.9  &  4.4  &  0.3 &  4.4   &  0.2  \\
        $\bar{\ell}_6$ & \textcolor{blue}{14.3} & 18.7 &  1.1  & 16.0  &  \text{$0.5 \pm 0.7$}  & \multicolumn{2}{c}{\textbf{}} \\
        \midrule
        & 
        & \multicolumn{2}{c}{\textbf{Exp.}} &  \multicolumn{2}{c}{\textbf{W}}  & \multicolumn{2}{c}{\textbf{}} \\
        \midrule
        $\lambda$       & \gtbblue{0.02} &   \multicolumn{2}{c}{} &  \multicolumn{2}{c}{0.023}   &  \multicolumn{2}{c}{} \\
        $\fpi$ ($\MeV$) & \gtbblue{101}   & \multicolumn{2}{c}{92}  &  \multicolumn{2}{c}{}   &  \multicolumn{2}{c}{}  \\
        \bottomrule
    \end{tabular}
    \caption{Comparison of low-energy constants \( \bar{\ell}_i \)  (see appendix \ref{chiPT} for definition) computed in this paper ({\bf GTB}) with analysis of experimental results: {\bf GL} \cite{GASSER1984142}, {\bf Bij} \cite{Bijnens:1994ie,Bijnens:1997vq,Bijnens:1998fm}, {\bf CGL}\cite{Colangelo:2001df}. The value of $\lambda=0.023$ is from the (tree-level) Weinberg model ({\bf W}) \eqref{lambdaW} and $\fpi\simeq 92\MeV$ is the standard value \cite{Workman:2022ynf}. There is reasonable agreement except for $\bar{\ell}_{1,3}$ that are quite difficult to determine.}
    \label{tab:LECs1}
\end{table}

\begin{table}[h]
    \centering
    \renewcommand{\arraystretch}{1.4}
    \begin{tabular}{
        c
        ||c
        >{\raggedleft\arraybackslash}p{1.5cm}
        @{\,$\pm\,$}
        S[table-format=1.4]
        >{\raggedleft\arraybackslash}p{1.5cm}
        @{\,$\pm\,$}
        S[table-format=1.4]
        c
    }
        \toprule
        & \textbf{GTB} 
        & \multicolumn{2}{c}{\textbf{GL}} 
        & \multicolumn{2}{c}{\textbf{PY}} 
        & \textbf{W} \\
        \midrule
        $a^{(0)}_0$               & \gtbblue{0.172}    & 0.220  & 0.005  & 0.230  & 0.010  & 0.16 \\
        $a^{(2)}_0$               & \gtbblue{-0.0366}  & -0.0444 & 0.0010 & -0.0422 & 0.0022 & -0.046 \\
        $b^{(0)}_0$               & \gtbblue{0.251}    & 0.280  & 0.001  & 0.268  & 0.010  & 0.18 \\
        $b^{(2)}_0$               & \gtbblue{-0.063}   & -0.080 & 0.001  & -0.071 & 0.004  & -0.092 \\
        $10^3 \times a^{(1)}_1$   & \gtbblue{32.1}     & 37.0   & 0.13   & 38.1   & 1.4    & 31 \\
        $10^3 \times b^{(1)}_1$   & \gtbblue{2.69}     & 5.67   & 0.13   & 4.75   & 0.16   & 0 \\
        $10^4 \times a^{(0)}_2$   & \gtbblue{12.9}     & 17.5   & 0.3    & 18.0   & 0.2    & 0 \\
        $10^4 \times a^{(2)}_2$   & \gtbblue{-1.1}     & 1.70   & 0.13   & 2.2    & 0.2    & 0 \\
        \midrule
        & 
        & \multicolumn{2}{c}{\textbf{Exp. fits}} & \multicolumn{3}{c}{\textbf{}} \\
        \midrule
        $\langle r^2 \rangle^\pi_{S}$ ($\mathrm{fm}^2$) & \gtbblue{0.55}  & 0.61 & 0.04 & \multicolumn{2}{c}{} & \\
        $\langle r^2 \rangle^\pi_{V}$ ($\mathrm{fm}^2$) & \gtbblue{0.441} & 0.439 & 0.0087 & \multicolumn{2}{c}{} & \\
        $\langle r^2 \rangle^\pi_{T}$ ($\mathrm{fm}^2$) & \gtbblue{0.146} & \multicolumn{4}{c}{} & \\
        \bottomrule
    \end{tabular}
    \caption{Comparison of scattering lengths and pion radii as computed here ({\bf GTB}) with experimental values from {\bf GL} \cite{GASSER1984142}, {\bf PY} \cite{Pelaez:2004vs} and the tree level Weinberg model ({\bf W}) \cite{PhysRevLett.17.616}.}
    \label{tab:LECs2}
\end{table}

\begin{figure}[h!]
 	\centering
    \IfFileExists{figures/Amandel.pdf}{
    \includegraphics[width=\textwidth]{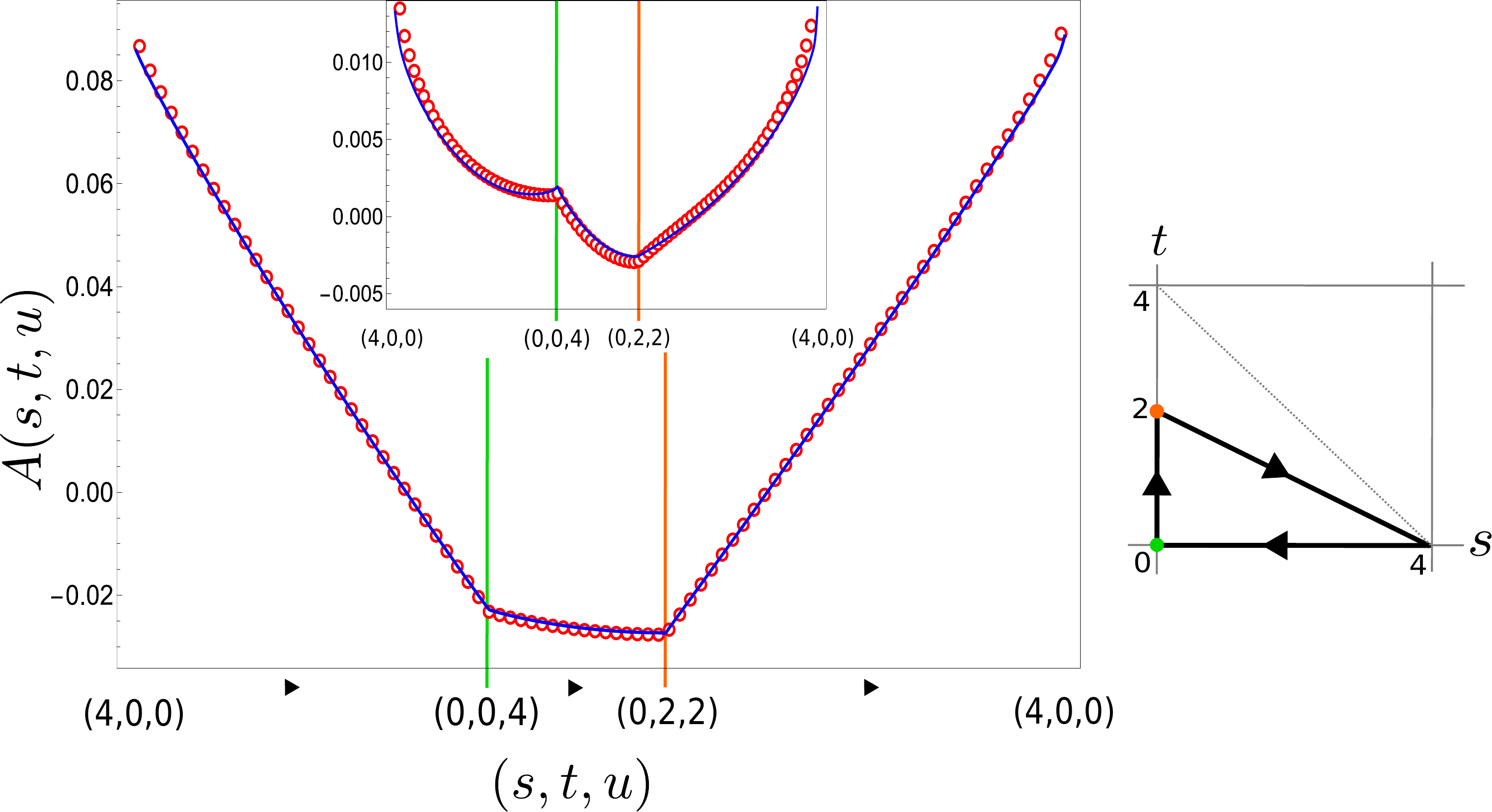}
}{
    \textbf{Figure is not available.}
}
 	\caption{We follow the value of the pion scattering amplitude $A(s,t,u)=A(s,u,t)$ along the path indicated on the right in the $(s,t)$ plane. The blue line is the prediction of $\chiPT$ using tree-level plus one-loop with the values of the constants extracted from our result as shown in table \ref{tab:LECs1}. The red circles are the results of the GTB converged solution. The inset depicts the same with the tree-level part removed.}
 	\label{A1loop}
\end{figure}

We can then plot the computed amplitude $A(s,t,u)$ and compare it with the $\chiPT$ results using the low energy coefficients we extracted. To simplify the plot we follow the value of the amplitude along a path in the Mandelstam triangle as depicted in fig. \ref{A1loop}.  The matching is a check of the accuracy of the results.

\subsection{Resonances}

\begin{table}[h!]
\centering
\renewcommand{\arraystretch}{1.3}
\begin{tabular}{l||c>{\raggedleft\arraybackslash}p{1.5cm}
        @{\,$\pm\,$}
        l
        c}
\toprule
 & \textbf{GTB} & \multicolumn{2}{c}{\textbf{\ \ \ \ PDG}} & \\
\midrule
$m_\rho$         & \gtbblue{758} & 775 & 0.23        & MeV \\
$\Gamma_\rho$    & \gtbblue{137} & 149.1 & 0.8  & MeV \\
$m_{\rho'}$      & \gtbblue{1514} & 1465 & 25     & MeV \\
$\Gamma_{\rho'}$ & \gtbblue{162} & 400 & 60  & MeV \\
$m_{f_2}$        & \gtbblue{1180} & 1275.4 & 0.8    & MeV \\
$\Gamma_{f_2}$   & \gtbblue{249} & 186.6 & 2.3 & MeV \\
\bottomrule
\end{tabular}
\caption{Masses and widths of the most prominent resonances. The results are similar and compatible with our previous work \cite{PhysRevLett.133.191601,PhysRevD.110.096001} and agree reasonably well with the known experimental results \cite{Workman:2022ynf}.}
\label{masses}
\end{table}

\begin{figure}[h!]
 	\centering	\includegraphics[width=0.7\textwidth]{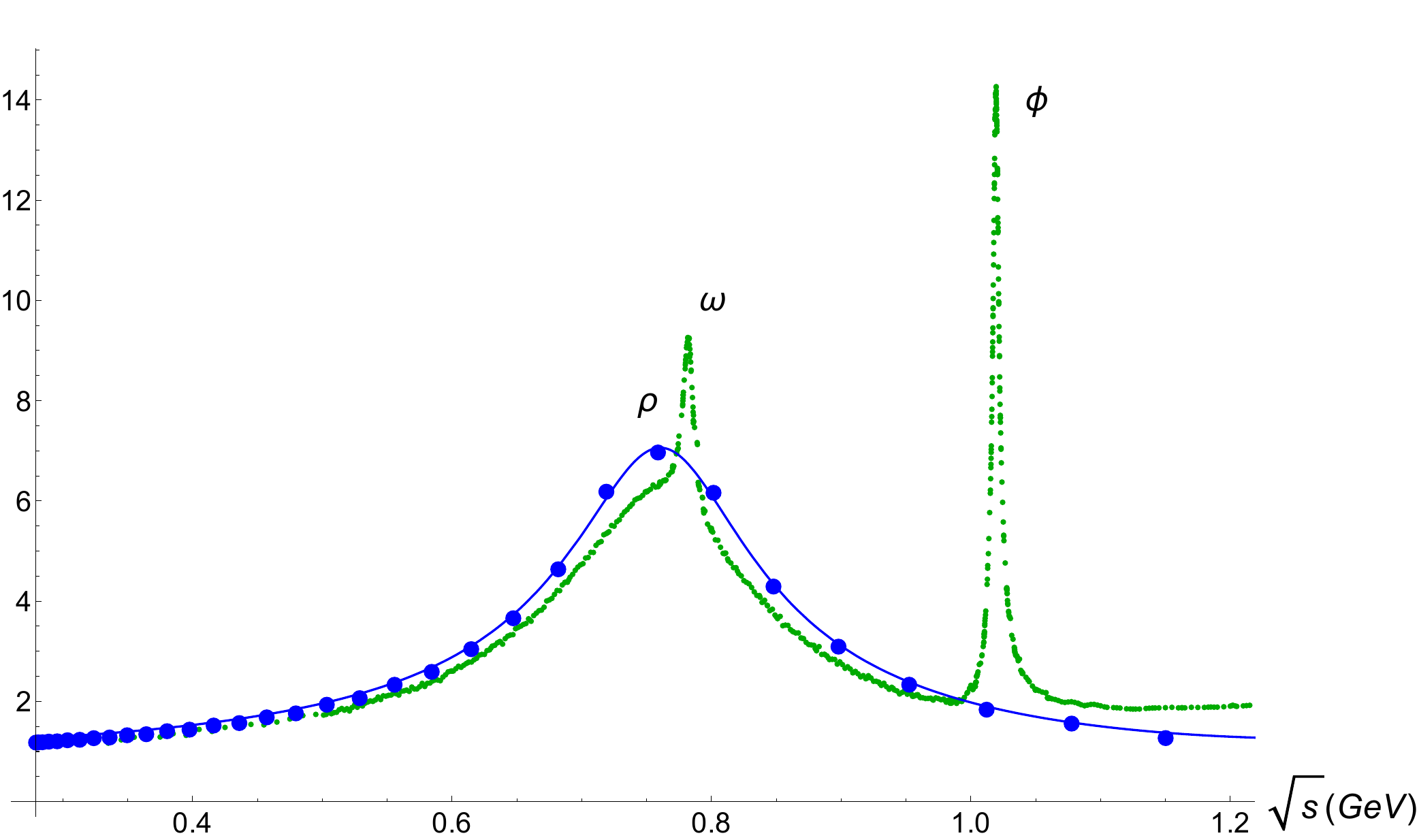}
 	\caption{In green hadronic R-ratio taken from \cite{Workman:2022ynf} and rescaled to match the vector form factor $|F^1_1|$ and compared with the present result (blue). The  $N_f=2$ vector current does not couple to the $\omega$ and $\phi$ resonances so we do not see those peaks. The agreement is reasonable and of interest in computations of the hadronic contributions to $g-2$ based on this method \cite{Zahed:2024haa} }
 	\label{Rratio}
 \end{figure}

As seen in fig. \ref{pw}, the partial waves and form factors display very clear resonances in the $P1$ and $D0$ channel whose width and masses can be extracted by fitting the form factors with a Breit-Wigner shape. The results are summarized in table \ref{masses}. Once again we would like to emphasize that no assumption was made on the number of resonances or any other property of the spectrum. The existence of these resonances is a result of the calculation. Finally, one can notice broad resonances in the $S0$ channel, the well-known $\sigma/f_0(500)$ whose nature is somewhat controversial \cite{Pelaez:2015qba} (and perhaps additional $f_0$ states) that we do not list in the table. Also, the $P1$ form factor is of interest due to its relation to the hadronic ratio $R$ \cite{Workman:2022ynf} that measures hadronic photo-production in $e^+e^-$ collisions. We compare the results of this paper with experiment in fig. \ref{Rratio}. The agreement is good although we are basically showing the $\rho$ meson dominance of the form factor. 

\section{Dependence on the input parameters}
\label{chichange}

\subsection{Form factor asymptotics}\label{sec:FF}
\begin{figure}[h!]
		\begin{subfigure}[h]{0.5\textwidth}
			\centering
			\includegraphics[width=\textwidth,trim={0cm 0cm 0cm 0cm},clip]{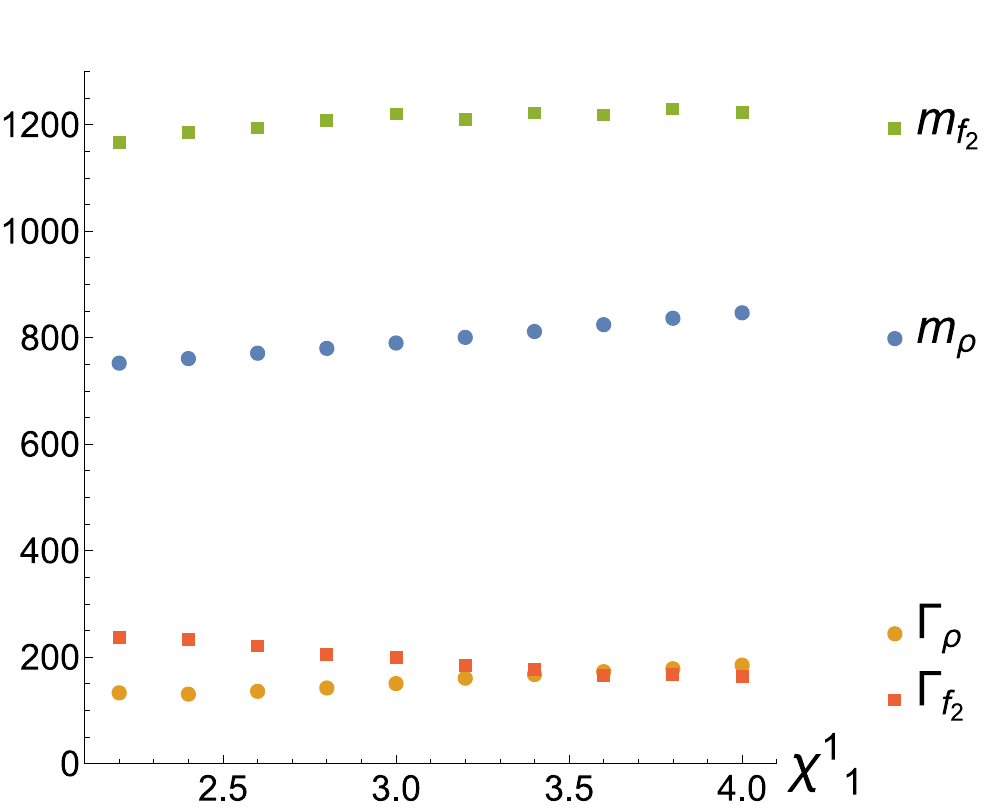}
			\caption{}
			\label{masswidth}
		\end{subfigure}
		\begin{subfigure}[h]{0.5\textwidth}
			\centering
			\includegraphics[width=\textwidth, trim={0cm 0cm 0cm 0cm},clip]{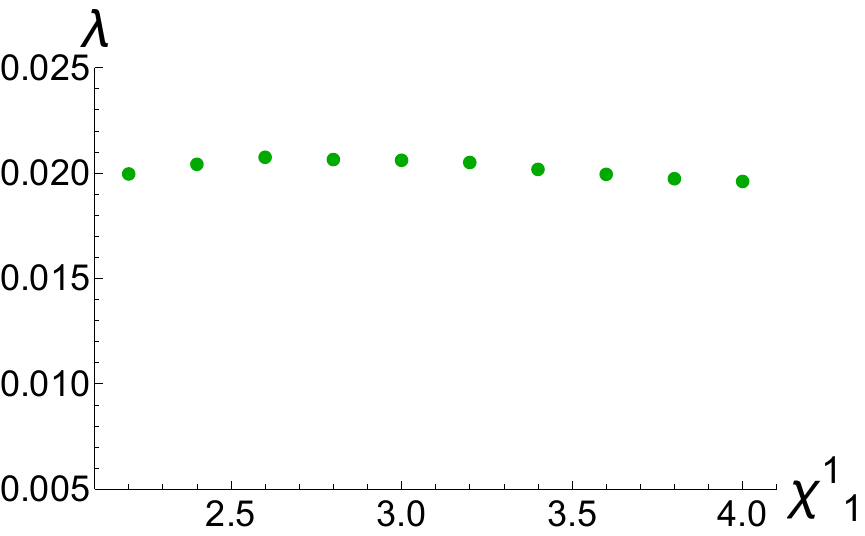}
			\caption{}
			\label{psff}
		\end{subfigure}
 	\caption{the $\rho,f_2$ meson mass and width (in MeV) \eqref{masswidth} and pion coupling $\lambda$ \eqref{psff} for form factor asymptotic factor $2.2\le \chi^1_1\le 4$.}
 \label{rhomass1}
 \end{figure}

\begin{figure}[h!]
 	\centering
    \IfFileExists{figures/FFps.pdf}{
    \includegraphics[width=\textwidth]{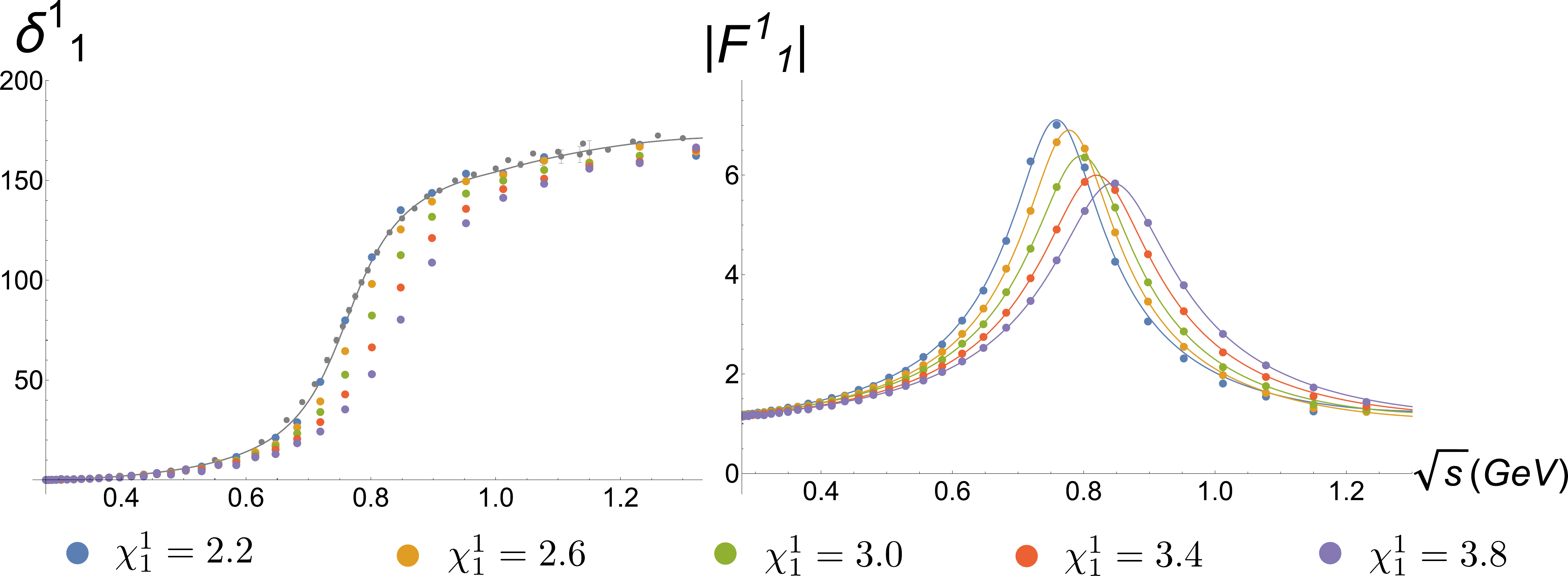}
}{
    \textbf{Figure is not available.}
}
 	\caption{The P1 phase shift $\delta^1_1$ and vector form factor $|F^1_1(s)|$ for form factor asymptotic factor $2.2\le \chi^1_1\le 3.8$}
 	\label{psffchi}
\end{figure}

\begin{figure}[h!]
		\begin{subfigure}[h]{0.5\textwidth}
			\centering
			\includegraphics[width=\textwidth,trim={0cm 0cm 0cm 0cm},clip]{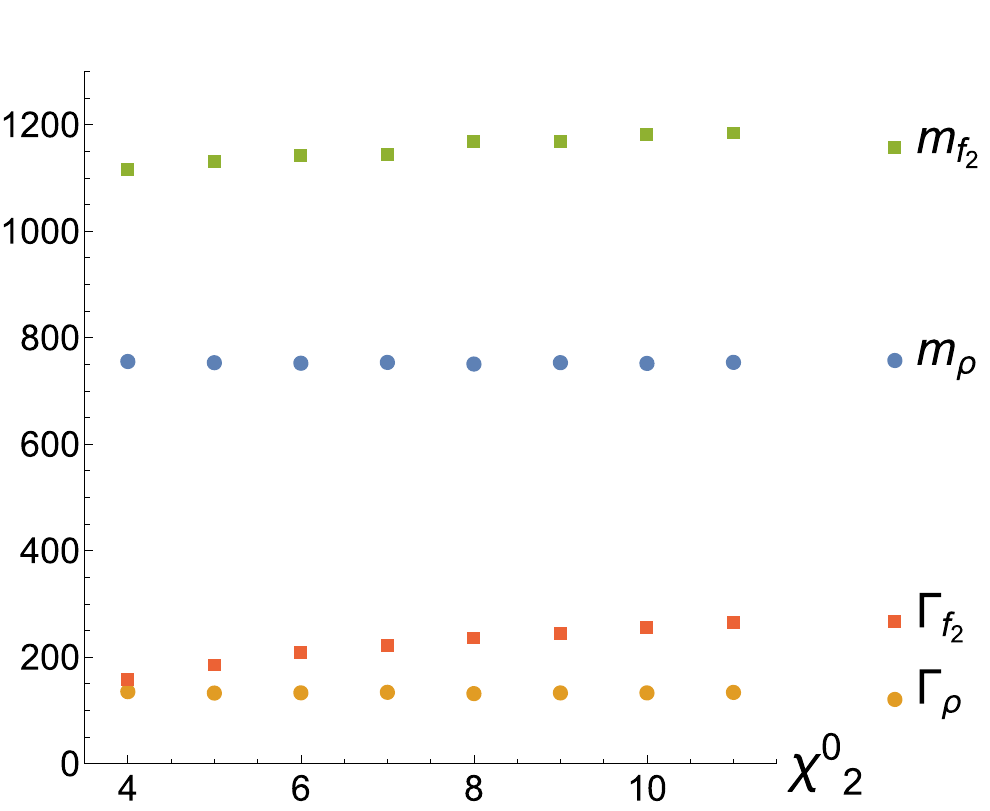}
			\caption{}
			\label{masswidth2}
		\end{subfigure}
		\begin{subfigure}[h]{0.5\textwidth}
			\centering
			\includegraphics[width=\textwidth, trim={0cm 0cm 0cm 0cm},clip]{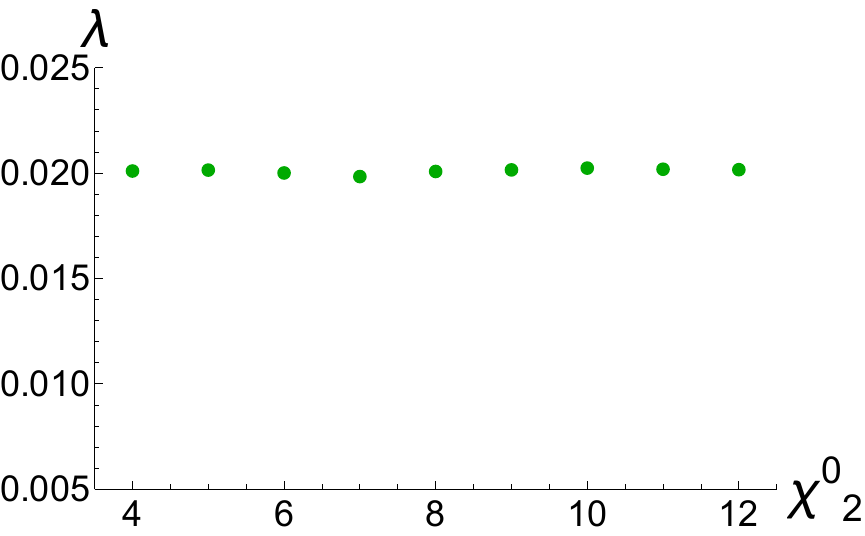}
			\caption{}
			\label{psff2}
		\end{subfigure}
 	\caption{the $\rho,f_2$ meson mass and width (in MeV) \eqref{masswidth2} and pion coupling $\lambda$ \eqref{psff2} for form factor asymptotic factor $4\le \chi^0_2\le 12$.}
 \label{rhomass2}
 \end{figure}

\begin{figure}[h!]
		\begin{subfigure}[h]{0.5\textwidth}
			\centering
			\includegraphics[width=\textwidth,trim={0cm 0cm 0cm 0cm},clip]{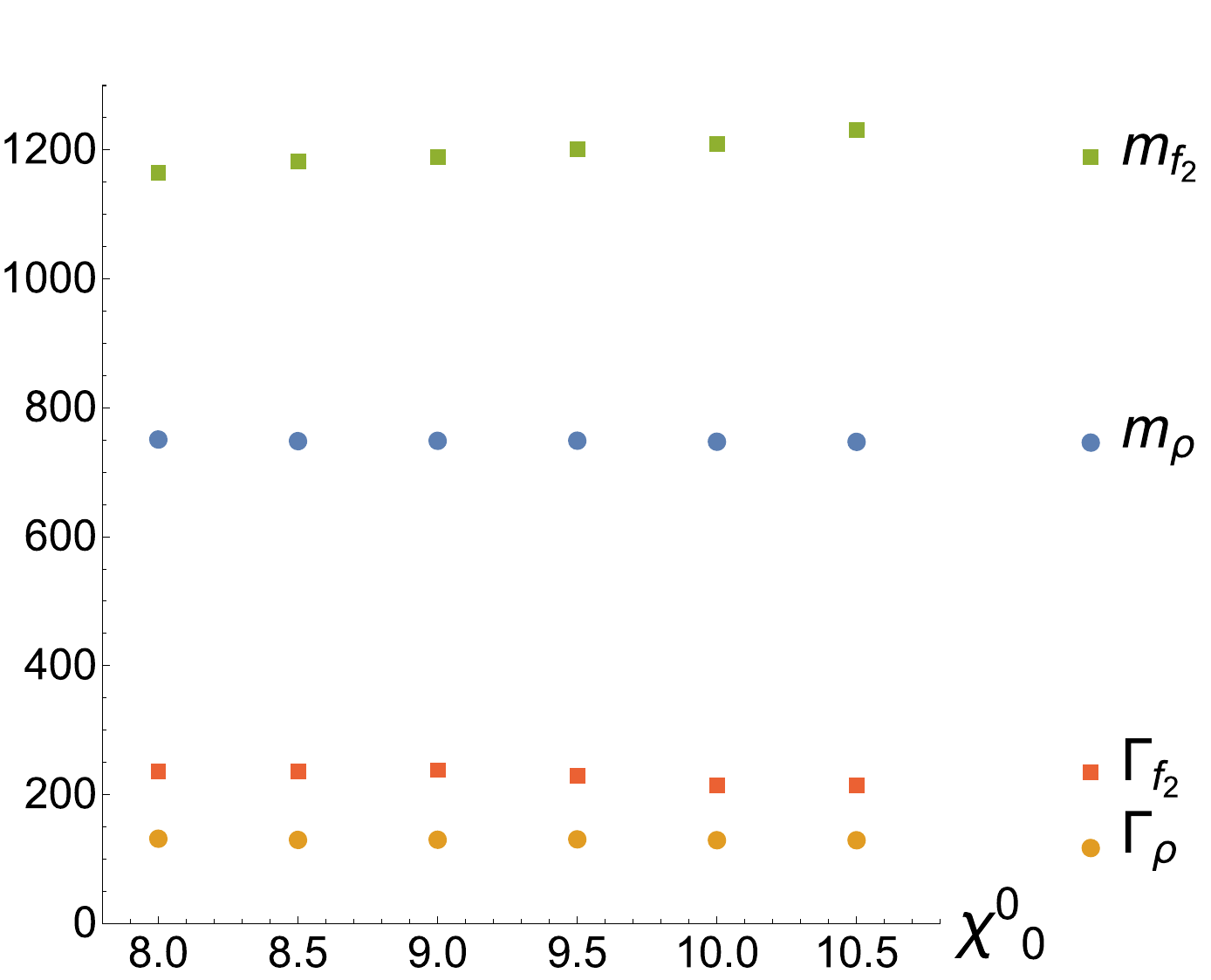}
			\caption{}
			\label{masswidth3}
		\end{subfigure}
		\begin{subfigure}[h]{0.5\textwidth}
			\centering
			\includegraphics[width=\textwidth, trim={0cm 0cm 0cm 0cm},clip]{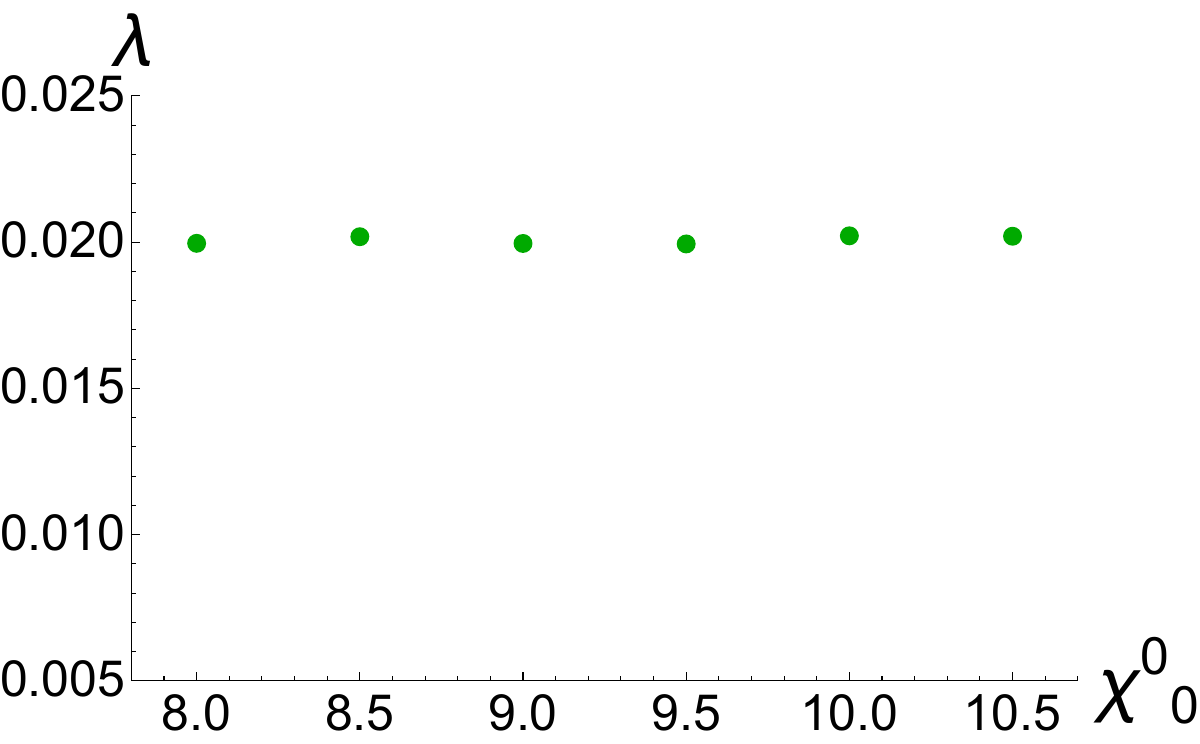}
			\caption{}
			\label{psff3}
		\end{subfigure}
 	\caption{the $\rho,f_2$ meson mass and width (in MeV) \eqref{masswidth3} and pion coupling $\lambda$ \eqref{psff3} for form factor asymptotic factor $8\le \chi^0_0\le 10.5$.}
 \label{rhomass3}
 \end{figure}

As mentioned before, an input of the procedure is the asymptotic value of the form factors. In the literature it has been argued, for example, that, at $2\GeV$ the P1 form factor can differ from the Brodsky-Lepage value \cite{osti_1447331,Pire:1996bc} by a factor up to 3 \cite{PhysRevLett.111.141802}. For that reason we explore how the $\rho$ mass and width, the $f_2$ mass and width change as a function of this parameter starting from minimal feasible value.  The results are in figures \ref{rhomass1}, \ref{psffchi}, \ref{rhomass2}, \ref{rhomass3}.

\subsection{Low energy tolerance}
\begin{figure}[h!]
 	\centering
    \includegraphics[width=\linewidth]{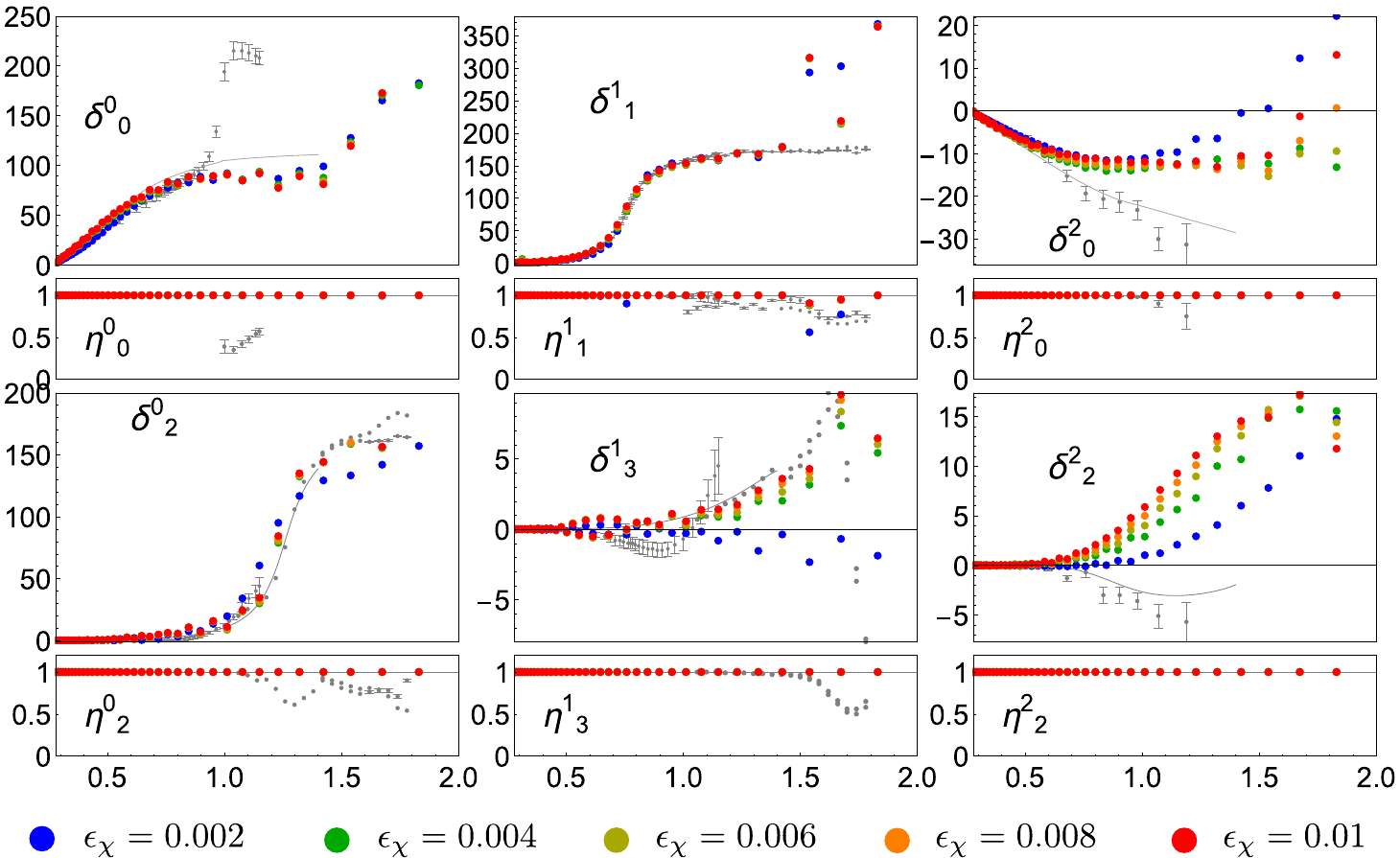}
 	\caption{Partial waves after 10 iteration with different low energy tolerance $\epsilon_{\chi}$}
 	\label{pwvchi}
 \end{figure}

For completeness we also study the dependence of the converged solution on the tolerance used to match the low energy partial waves \cite{PhysRevLett.133.191601,PhysRevD.110.096001}. At low energy we choose a small set of points $s_j$ in the very low energy unphysical region $0<s_j\le 2$, and require that
\begin{equation}\label{chiralconstraints} 
\begin{aligned}
||f^0_0(s_j)-R^{\text{tree}}_{01}(s_j)f^1_1(s_j)||&\le \epsilon^{\chi},\\
||f^2_0(s_j)-R^{\text{tree}}_{21}(s_j)f^1_1(s_j)||&\le \epsilon^{\chi},\\
R^{\text{tree}}_{01}=\frac{f^0_{0,\text{tree}}}{f^1_{1,\text{tree}}},\;\;R^{\text{tree}}_{21}=&\frac{f^2_{0,\text{tree}}}{f^1_{1,\text{tree}}}
\end{aligned}
\end{equation}
with some  tolerance $\epsilon^\chi$. The values referred as ``tree" are from the tree level Weinberg model \eqref{h5}. The results for various values of the tolerance $\epsilon^{\chi}$ are displayed in fig.\ref{pwvchi}. Once again, the Watsonian iteration removes the dependence on this parameter except somewhat for the $I=2$ channels (and also spin 3) where we do not have UV input.

\clearpage

\section{Thermodynamics of a dilute gas of pions, second virial coefficients and density of pion pairs}

The computation of the two particle elastic S-matrix allows us to study the thermodynamics properties of a pion gas in the low density and low energy region \cite{Huang,PhysRev.187.345}. A dilute gas of bosons is such that each single particle state $\epsilon_\alpha$ has a low occupation number $n_\alpha\ll 1$. In that case, the virial expansion determines the partition function in terms of the single particle partition function with corrections from two particle elastic scattering that we can compute from our results in a similar way as done in \cite{VENUGOPALAN1992718} based on experimental phase shifts. This is part of a more general program of computing the thermodynamics of relativistic gases based on the scattering matrix \cite{PhysRev.187.345}. See also \cite{baratella2024thermodynamicssmatrixreloadedapplications} for a recent discussion in the context of QCD strings. Briefly, the grand canonical partition function $\Xi$ can be expanded in powers of the fugacity $z=e^{\beta\mu}$ as
\beq
\beta P = \lim_{V\rightarrow \infty}\frac{1}{V} \ln \Xi = \sum_{n=1}^\infty b_n(T) z^n
\eeq
where \eg\ from \cite{VENUGOPALAN1992718} we have
\begin{subequations}
\beqa
b_1 &=& 3\int \frac{d^3 p}{(2\pi\hbar)^3}\, e^{-\beta\sqrt{p^2+\mpi^2}} = \frac{3\mpi^2}{2\pi^2\hbar^3\beta}\, K_2(\beta\mpi)\\
b_2 &=& \frac{1}{2\pi^3\beta} \int_{2m_{\pi}}^\infty dM\, M^2 K_2(\beta M) \sum_{I\ell}{}^{'} (2I+1)(2\ell+1) \frac{\partial \delta^I_\ell}{\partial M}
\eeqa
\end{subequations}
where $K_2(x)$ is a standard modified Bessel function and $\delta^I_\ell$ are the scattering phase-shifts. If pions are created by a transition from an unconfined phase, the number of pions will maximize the 
entropy and therefore the chemical potential would be zero. In general one would also expect $\mu=0$ since the number of pions is not conserved, but inelastic scattering is highly suppressed in the dilute gas and low energy approximations so an effective $\mu$ can appear. Presumably we can even have Bose-Einstein condensation. However this is only on the assumption that electromagnetic and weak interactions are disconnected so that pions are stable (as in lattice calculations for example).

\begin{figure}[h!]
 	\centering
    \IfFileExists{figures/piongas.pdf}{
    \includegraphics[width=0.9\linewidth]{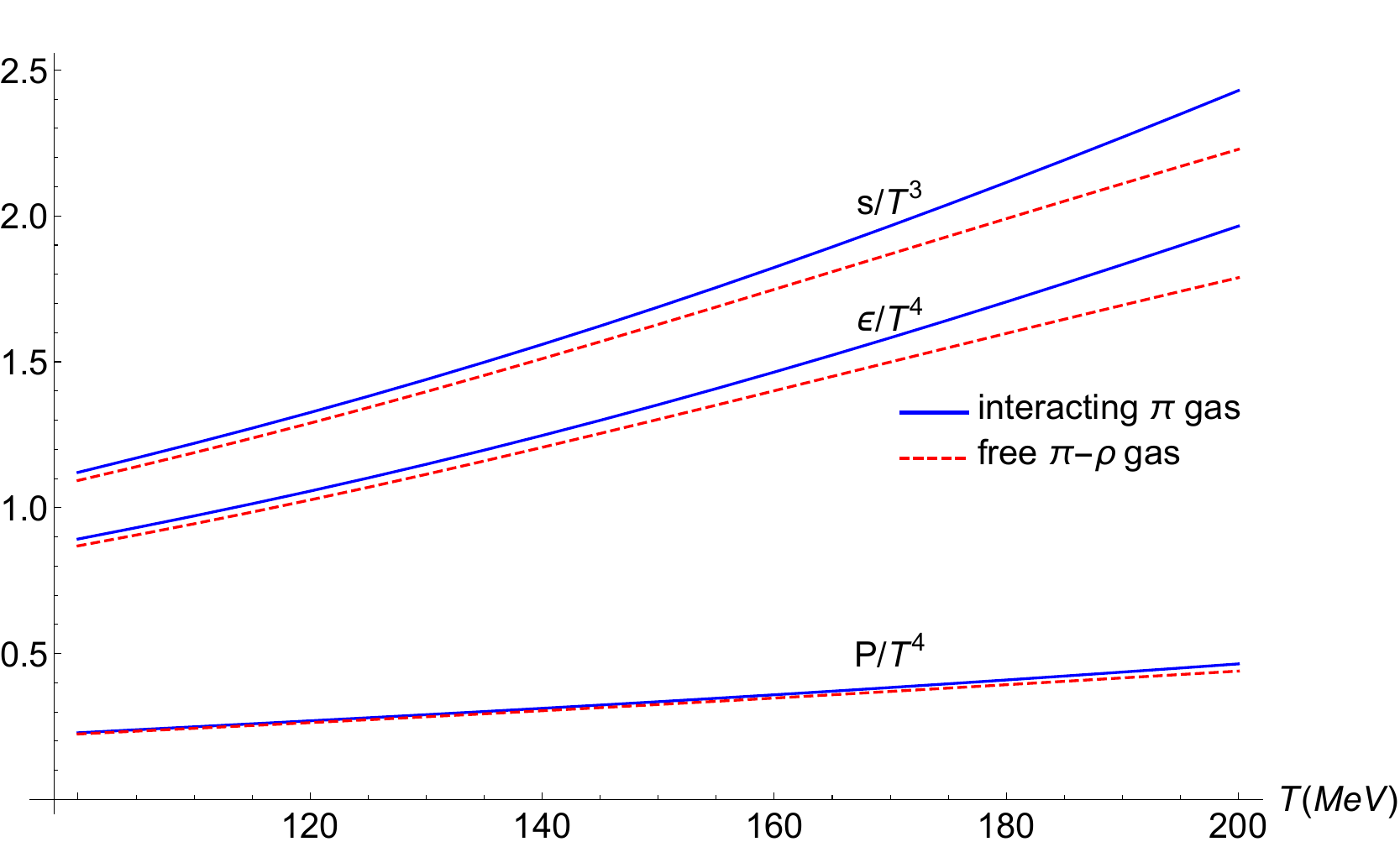}
}{
    \textbf{Figure is not available.}
}
 	\caption{Thermodynamic quantities (pressure $P$, energy density $\epsilon$, entropy density $s$) of a dilute interacting pion gas as a function of temperature for $\mu=0$, computed from the phase shifts we obtained (fig. \ref{pw}). We compare with a free gas of pions and $\rho$ mesons. As discussed in \cite{VENUGOPALAN1992718} the results are similar.}
 	\label{PT}
 \end{figure}

 \begin{figure}[h!]
 	\centering
    \IfFileExists{figures/Nosssplot.pdf}{
    \includegraphics[width=\linewidth]{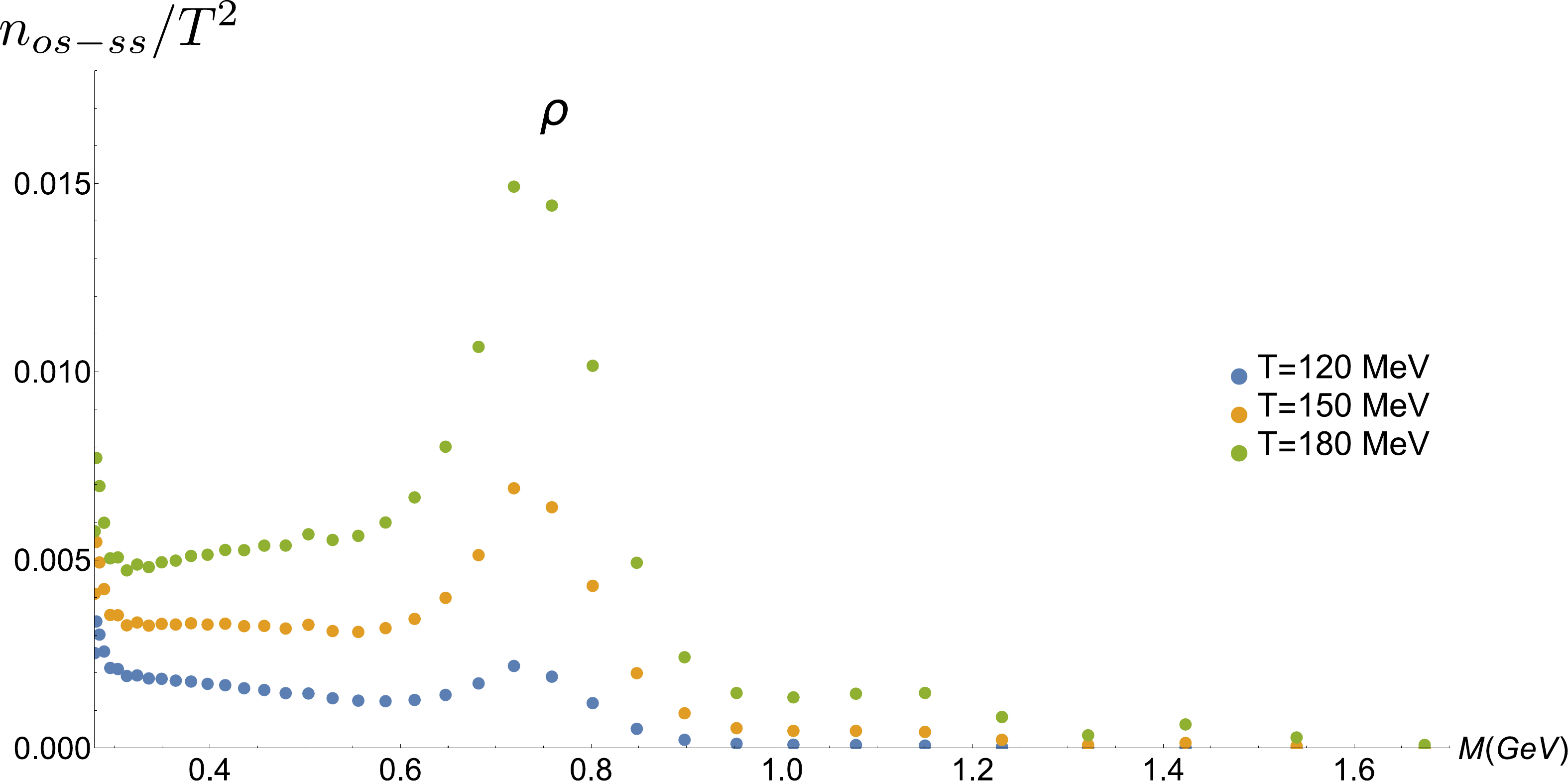}
}{
    \textbf{Figure is not available.}
}
 	\caption{From \eqref{nosss}, $n_{os-ss}/T^2$, \ie\ density of pairs of pions with opposite sign minus same sign distributed over the invariant mass $M$ and normalized with $T^2$. The results are for a dilute gas of pions using the phase shifts obtained in this paper. The chemical potential is taken to be $\mu=0$.}
 	\label{NosminusNss}
 \end{figure}

From here we can compute the pressure as a function of temperature (for $\mu=0$)
\begin{equation}
    \begin{aligned}
P_{\text{int}} = T b_2, \;
\epsilon_{\text{int}} = -\frac{\partial b_2}{\partial \beta},\;
n_{\text{int}} = 2 b_2, \;
s_{\text{int}} = \left[ b_2 \left(1 - 2 \mu \beta \right) - \beta \frac{\partial b_2}{\partial \beta} \right]
\end{aligned}
\end{equation}
where ``int" denotes the interaction correction to the ideal gas expressions.
The results are displayed in fig.\ref{PT} and, as expected are similar to those in \cite{VENUGOPALAN1992718}. 

\subsection{Pairs of pions with opposite and same sign of electric charge}
A quantity that is of interest for pions produced in high energy nuclear collisions\footnote{We are grateful to Fuqiang Wang for a discussion on this topic.} \cite{STAR:2020gky}, when the quark gluon plasma cools down, is the difference in the pairs of pions with opposite (electric) charge $N_{\mbox{os}}$ minus the number of pairs with the same sign $N_{\mbox{ss}}$. If we consider that quantity in a gas at equilibrium, as we do here, then we define
\beq
N_{\mbox{os}}-N_{\mbox{ss}} = 2 \langle\langle N_+ N_-\rangle\rangle - \langle\langle N_+ N_+\rangle\rangle - \langle\langle N_- N_-\rangle\rangle 
\eeq
where the brackets $\langle\langle\rangle\rangle$ indicate thermal average.
 For a dilute gas of free pions this difference vanishes since \eg\ $\langle\langle N_+ N_-\rangle\rangle =  \langle\langle N_+\rangle\rangle \langle\langle N_-\rangle\rangle$  and $\langle\langle N_+\rangle\rangle = \langle\langle N_-\rangle\rangle$. In an interacting gas, this is no longer true, the eigenstates of the Hamiltonian that enter the partition function are not eigenstates of $N_\pm$.  Suppose we want to compute the number of pairs with opposite sign
\beq
N_{OS} = \langle\langle N_+ N_- \rangle\rangle = \frac{1}{\Xi} \sum_{\nu} N_+ N_- e^{-\beta E_\nu + \beta \mu N_\nu}  
\eeq
Expanding in powers of the fugacity it is clear we need at least two particles to have a pair so we get, at order $z^2$:
\beqa
N_{OS}-N_{OS}^{free} &=& z^2  \sum_{\nu, N_\nu=2}  N_+ N_- e^{-\beta E_\nu} \\
       &=& V z^2 \int dM \frac{M^2}{2\pi^2\beta}\, K_2(\beta M) \sum_{\ell,I,I_z} (2\ell+1)  \bra{II_z} N_+ N_- \ket{II_z} \frac{1}{\pi} \frac{\partial\delta^{I}_\ell}{\partial M} \nonumber
\eeqa
In the sum over $I_z$ only $I_z=0$ survives. Now we have
\beq
\bra{I 0} N_+N_- \ket{I0}  = |\bra{I0} +-\rangle|^2=\left\{\begin{array}{cc} =\frac{1}{3} , \ \ I=0  \\ \\ =\frac{1}{2} , \ \ I=1  \\ \\ = \frac{1}{6}, \ \ I=2\\\end{array}\right.
\eeq
whereas for same sign we have $I_z=2$ and then $I=2$. Putting everything together we find 
\beqa
N_{OS}-N_{SS} 
&=& V z^2 \int_{2m_{\pi}}^{\infty} dM \frac{M^2}{2\pi^2\beta}\, K_2(\beta M) \sum_{\ell}{}' (2\ell+1)  \frac{1}{\pi}\frac{\partial}{\partial M}\left( \frac{1}{3}\delta^0_\ell +\frac{1}{2}\delta^1_\ell-\frac{5}{6}\delta^2_\ell\right)\nonumber
\eeqa
where $\sum{}'$ indicates that the sum is over even or odd $\ell$ for even or odd isospin. Still using the distribution in $M$ ($M^2=(p_1+p_2)^2$) but introducing $s=M^2$ we get:
\beq
\begin{aligned}
N_{OS}-N_{SS} =&V z^2 \int_{2m_{\pi}}^{\infty} dM\;n_{os-ss}(M)\\
= &V z^2 \int_{2m_{\pi}}^{\infty} dM \frac{s^{\frac{3}{2}}}{\pi^3\beta} K_2(\beta\sqrt{s}) \frac{\partial}{\partial s} \left(\frac{1}{3}\delta_{S0}+\frac{3}{2}\delta_{P1}-\frac{5}{6}\delta_{S2} +\frac{5}{3}\delta_{D0}\right)\label{nosss}
\end{aligned}
\eeq 
 The results are depicted in fig. \ref{NosminusNss} for the case of zero chemical potential $z=1$. The $\rho$ resonance peak coming from $\frac{\partial \delta_1^1}{\partial M}$ is evident. It can be interpreted as an increase in the density of states near the $\rho$ mass or as $\rho$ mesons being created and destroyed leading in both cases to an extra density of pairs of opposite sign.


\section{Conclusions} 

The goal of the Gauge Theory Bootstrap is to completely determine the dynamics of asymptotically free gauge theories (QCD) in the strongly coupled regime where perturbative QCD does not apply and neither does the low energy effective pion Lagrangian. An important point is that the only stable particles are pions (assuming no other interactions are present and ignoring nucleons)\ and therefore pions are the only asymptotic states in that energy region. The idea is then to constrain the dynamics using input from the very low energy Lagrangian (using $\mpi$) and from high energy using sum rules and asymptotic form factors. Notice that the leading asymptotic form factors are basically controlled by the pion radius that is determined by $\fpi$. Therefore we also need, from low energy at least the order of magnitude of $\fpi$. Given those and the gauge theory parameters we expect to find a unique pion scattering matrix, form factors, spectral densities, etc. In this paper we introduce an iterative procedure that, given the UV and IR information, converges to a unique answer which we take to be the result of the computation.  By maximizing and minimizing various physical quantities (like couplings, scattering lengths, etc.) we show that there is a small space of qualitatively similar solutions around the converged one providing an estimate on the error of the computation. Another source of error is the fact that the asymptotic values of the form factors are not predicted from QCD very precisely. Finally there is a systematic error due to the fact that we take the converged solution as the best estimate of the physical solution. This could be improved, for example,  by including more operators in each channel. We expect to explore this more in future work.

\section{Acknowledgements}
We are grateful to Lance Dixon, Nick Manton, Ian Moult, Massimiliano Procura and Arkady Vainshtein for many interesting discussions on the topics of this paper. We would like to thank Slava Rychkov for comments on the previous version of this paper.
 Y.H. is grateful to CERN for hospitality while this work was being completed. 
This work was supported in part by DOE through the QuantiSED Fermilab consortium.

\appendix
\renewcommand{\theequation}{\Alph{section}.\arabic{equation}}
\setcounter{equation}{0} 

\section{Chiral  Lagrangian and one-loop amplitude}\label{chiPT}

The very low energy description of pion dynamics is given by the non-linear sigma model:
\beq
\cL_2 =  \frac{F^2}{4}\tr[D_\mu U D^\mu U^\dagger] + \frac{F^2}{4} \tr(\chi U^\dagger+U \chi^\dagger)
\eeq
where $U=\exp(i\pi^a\tau_a/F)$ and covariant derivatives include external gauge fields as sources.  
Also $\chi$ is a source whose expectation value determines the pion mass\footnote{ It is conventional to call this source $\chi$ as we do in this appendix. It should not be confused with $\chi^I_\ell$ in the main text that represents the factors in the form factor asymptotics.}. We refer to \cite{GASSER198765,GASSER1984142} for the notation and main summary of the chiral perturbation theory approach. In particular, in this appendix we follow \cite{GASSER1984142} and denote as $F$, $M$ the parameters of the effective action whereas $\fpi$, $\mpi$ denote the physical values of the pion decay constant and mass (see \eqref{massM} below).  At this order,  the pion scattering amplitude is
\beq\label{h5}
A_{\text{tree}}(s,t,u) =  \frac{s-M^2}{8\pi^2\,F^2}
\eeq
This is the $t\leftrightarrow u$ symmetric part. The full amplitude for $\pi_a(p_1) +\pi_b(p_2) \rightarrow \pi_c(p_3) +\pi_d(p_4)$ is written as
\beq
T_{ab,cd} = A(s,t,u) \delta_{ab}\delta_{cd} +A(t,s,u) \delta_{ac}\delta_{bd} +A(u,t,s) \delta_{ad}\delta_{bc}
\eeq
where $a,b,c,d$ are isospin one indices.
 The higher energy corrections were studied systematically by Gasser and Leutwyler in the same work \cite{GASSER198765} leading to the definition of a number of low energy (or Wilson) coefficients $\bar{\ell}_j$ through the Lagrangian \cite{GASSER198765}:
\beqa\label{L4}
\cL_4 &=& \frac{l_1}{4}\left\{\tr[D_\mu U (D^\mu U)^\dagger]\right\}^2 + \frac{l_2}{4}\tr[D_\mu U(D_\nu U)^\dagger]\tr[D^\mu U(D^\nu U)^\dagger] \nonumber\\
&& +\frac{l_3}{16}[\tr(\chi U^\dagger+U\chi^\dagger)]^2 +\frac{l_4}{4}\tr[D_\mu U(D^\mu\chi)^\dagger+D_\mu\chi(D^\mu U)^\dagger] \nonumber\\
&& +l_5\left[\tr(f_{\mu\nu}^R U f_L^{\mu\nu} U^\dagger) -\half \tr(f_{\mu\nu}^L f^{\mu\nu}_L +f^R_{\mu\nu}f^{\mu\nu}_R) \right] \nonumber\\
&& +i\frac{l_6}{2}\tr[f_{\mu\nu}^R D^\mu U (D^\nu U)^\dagger + f_{\mu\nu}^L (D^\mu U)^\dagger D^\nu U ] \nonumber\\
&& -\frac{l_7}{16} [\tr(\chi U^\dagger - U \chi^\dagger)]^2 
\eeqa  
At one loop one introduces the renormalized constants $\bar{\ell}_j$ defined at a renormalization scale $\mu=\mpi$ \cite{GASSER1984142}. It also gives the physical $\mpi$, $\fpi$  as \cite{GASSER1984142} 
\beqa\label{massM}
\mpi^2 &=& M^2\left\{1-\frac{M^2}{32\pi^2 F^2} \bar{\ell}_3 +\cO(M^4)\right\} \\
\fpi   &=& F  \left\{1+\frac{M^2}{16\pi^2 F^2} \bar{\ell}_4 +\cO(M^4)\right\}
\eeqa
Although all this is well known, we include it here to give the definition of $\bar{\ell}_j$ and as a reference point for interpreting our results in comparison with other work. Notice, however, that our main focus is the intermediate energy region where pions are strongly coupled, namely where this Lagrangian does not apply. 
In chiral perturbation theory the low energy constants $\bar{\ell}_i$ are determined from fitting with experiments but in the Gauge Theory Bootstrap approach they are computed from QCD (under certain assumptions). 

One way to compute the $\bar{\ell}_j$'s is to match the amplitude we obtain in GTB to the tree-level plus one-loop result from \cite{GASSER1984142} (replacing $M$ by $\mpi$ using \eqref{massM}):
 \beq
 A(s,t,u) = \frac{s-m_{\pi}^2}{8\pi^2F^2}- \frac{\bar{\ell}_3}{256\pi^4 F^4} + \frac{1}{768\pi^4F^4}(A_1(s,t,u)+A_2(s,t,u)) + \cO(p^6)
 \eeq
 where 
 \begin{subequations}
     \beqa
 \!\!\!\!\!\!\!\!A_1(s,t,u) =&&\!\!\!\!\!\!\!\! 3(s^2-m_{\pi}^4)I_f(s)   + (t(t - u) - 2tm_{\pi}^2 + 4um_{\pi}^2 - 2m_{\pi}^4)I_f(t)  \\ 
 && + (u(u - t) - 2um_{\pi}^2 + 4tm_{\pi}^2 - 2m_{\pi}^4)I_f(u) \non \\
 \!\!\!\!\!\!\!\!A_2(s,t,u) =&&\!\!\!\!\!\!\!\! 2\left(\bar{\ell}_1 - \frac{4}{3}\right)(s - 2m_{\pi}^2)^2 + \left(\bar{\ell}_2 - \frac{5}{6}\right)(s^2 + (t - u)^2) - 12 sm_{\pi}^2 + 15m_{\pi}^4 \nonumber
 \eeqa
 \end{subequations}
 with
 \beq
 I_f(s) = \beta(s)\ln\left(\frac{\beta(s) - 1}{\beta(s) + 1}\right)+ 2,\;\;\;\;
 \beta(s) = \sqrt{1-\frac{4m_{\pi}^2}{s}}
 \eeq
 For $0<s<4m^2_{\pi}$ the expression
 \beq
 I_f(s) = 2-2\,\sqrt{\frac{4m_{\pi}^2}{s}-1}\ \mathrm{arccot}\!\left(\sqrt{\frac{4m_{\pi}^2}{s}-1}\right)
 \eeq
 is also useful.
 Introducing $\lambda$ as defined in \eqref{lambdadef1}, we obtain (we now set $\mpi=1$ as in most of the paper):
\beq\label{lambdadef2}
\lambda = \frac{1}{32\pi F^2} \left( 1 + \frac{1}{96 F^2}(1-\sqrt{2} \arctan(2\sqrt{2}) ) + \frac{1}{39 F^2} (\bar{\ell}_1+\bar{\ell}_2) \right) + \cO\left(\frac{1}{F^6}\right)
\eeq 
We also get:
\beq
A(s,t,u) = \frac{4}{\pi}\lambda (s-1) + \frac{4\lambda^2}{3\pi^2}\left[A_1(s,t,u)+A_2(s,t,u)-3(s-1)(\bar{A}_1+\bar{A}_2-3\bar{\ell}_3)-3\bar{\ell}_3\right] \nonumber
\eeq
 where $\bar{A}_{1,2}=A_{1,2}(\frac{4}{3},\frac{4}{3},\frac{4}{3})$.
 In particular notice the dependence on the coefficients $\bar{\ell}_{1,2,3}$:
 \beq\label{A1loopB} 
\begin{aligned}
&A(s,t,u;\bar{\ell}_1,\bar{\ell}_2,\bar{\ell}_3) =  A(s,t,u;0,0,0)   \\
     &+\frac{4\lambda^2}{3\pi^2}\left[2\bar{\ell}_1 (s - 2)^2 + \bar{\ell}_2 (s^2 + (t - u)^2) -\frac{8}{3}(s-1)(\bar{\ell}_1+2\bar{\ell}_2-\frac{27}{8}\bar{\ell}_3)-3\bar{\ell}_3\right]
\end{aligned}
 \eeq
 The physical value of $\fpi$ is related to $\lambda$ through
 \beq\label{fpicalc}
 \frac{1}{\fpi^2} =32 \pi  \lambda -\frac{32}{9} \lambda ^2 \left(8 \bar{l}_1+16 \bar{l}_2-27 \bar{l}_3+36 \bar{l}_4+33-33
   \sqrt{2} \tan ^{-1}\left(2 \sqrt{2}\right)\right)+\mathcal{O}(\lambda^3)
 \eeq
 It is interesting to study the amplitude in the Mandelstam triangle defined by $s>0$, $t>0$, $u>0$. Since $A(s,t,u)$ is symmetric under the interchange $t\leftrightarrow u$, we only need the subregion $t<u$. To simplify the plots we just consider the amplitude along the boundary of that triangle, as depicted in figure \ref{A1loop}. By matching the results of the computation with \eqref{A1loopB} we determine the values of $\bar{\ell}_{1,2,3}$. Further we can write the form factors near $s=0$ and at lowest non-trivial order as:
 \begin{subequations}\label{l46}
      \begin{eqnarray}
 F^{(0)}_0(s)&=&1+\frac{2\lambda}{\pi}s\left(\bar{\ell}_4-\frac{13}{12}\right)+\mathcal{O}(\lambda^2)\\
     F^{(1)}_1(s)&=&1+\frac{\lambda}{3\pi}s\left(\bar{\ell}_6-1\right)+\mathcal{O}(\lambda^2)
 \end{eqnarray}
 \end{subequations}
and use them to the determine $\bar{\ell}_{4,6}$. These expressions are used in the main text to obtain the results in table \ref{tab:LECs2}.
As a check and to get an idea of the errors in our computation, we can use the values in table \ref{tab:LECs2} to recompute the scattering lengths
from \cite{GASSER1984142}:
\begin{eqnarray}
         a^{(0)}_0&=&7 \lambda +\frac{\lambda ^2}{9 \pi } \left[64 \bar{\ell}_1+128 \bar{\ell}_2+144 \bar{\ell}_3+84+231 \sqrt{2} \tan ^{-1}(2
   \sqrt{2})\right]+\mathcal{O}(\lambda^3)\nonumber\\
   a^{(2)}_0&=&-2 \lambda +\frac{2\lambda ^2}{9 \pi } \left[32 \bar{\ell}_1+64 \bar{\ell}_2-36 \bar{\ell}_3+42-33 \sqrt{2} \tan ^{-1}(2 \sqrt{2})\right]+\mathcal{O}(\lambda^3)\nonumber\\
   a^{(1)}_1&=&\frac{4 \lambda }{3}+\frac{2 \lambda ^2}{27 \pi }\left[-64 \bar{\ell}_1+16 \bar{\ell}_2+54 \bar{\ell}_3-131+66 \sqrt{2} \tan ^{-1}(2
   \sqrt{2})\right]+\mathcal{O}(\lambda^3)\nonumber\\
   a_2^{(0)} &=& \frac{32\lambda^2}{45\pi}\left[\bar{\ell}_1+4\bar{\ell}_2-\frac{53}{8}\right]+\mathcal{O}(\lambda^3) \nonumber\\
a_2^{(2)} &=& \frac{32\lambda^2}{45\pi}\left[\bar{\ell}_1+\bar{\ell}_2-\frac{103}{40}\right]+\mathcal{O}(\lambda^3)\label{larelation}
\end{eqnarray}
and compare with the values in table \ref{tab:LECs2} extracted directly from the partial waves. See table \ref{arecompute} for the comparison.

\begin{table}[h!]
\centering
\renewcommand{\arraystretch}{1.3}
\begin{tabular}{c||cc}
\toprule
 & \textbf{GTB} & \textbf{GTB 2}\\
\midrule
$a^{(0)}_0$                 & \gtbblue{0.172} & 0.170\\
$a^{(2)}_0$                 & \gtbblue{-0.0366} & -0.0363\\
$10^3\times a^{(1)}_1$      & \gtbblue{32.1} & 29.7 \\
$10^4\times a^{(0)}_2$      & \gtbblue{12.9} & 15.0\\
$10^4\times a^{(2)}_2$      & \gtbblue{-1.1} & 4\\
\bottomrule
\end{tabular}
\caption{The scattering lengths extracted directly from the partial waves ({\bf GTB}) comparing with the ones calculated from our low energy constants $\bar{\ell}_i$ computation using \eqref{larelation} ({\bf GTB 2}). They are consistent except the $I=2$ scattering length that presently seems difficult to compute in this approach.}
\label{arecompute}
\end{table}


\section{Parameterization of the spectral density}\label{rhopara}

In this appendix we describe the parameterizations used for the spectral density, designed to satisfy the correct low energy behavior as described in section \ref{rholowE}. Consider, for example the vector current with the behavior \eqref{vecrho} at lowest order in $\chiPT$  \cite{GASSER198765} in our conventions. Setting $m_{\pi}=1$ we obtain, for $s\in\mathbb{R}_{\ge 4}$:
\beq
\rho^1_1(s) = \frac{1}{(2\pi)^4}\frac{s}{24\pi} \left(1-\frac{4}{s}\right)^{\frac{3}{2}} 
= \frac{1}{(2\pi)^4}\frac{1}{48\pi} (s-4)^{\frac{3}{2}} \left[ 1 -\frac{1}{8} (s-4) + \frac{3}{128} (s-4)^2 +\ldots \right]
\eeq
This suggest parameterizing a general $\rho$, in the region $4\le s\le s_0$ as
\beq
\rho^1_1(s) =  \frac{1}{(2\pi)^4}\frac{1}{48\pi} (s-4)^{\frac{3}{2}} \left[ B + \sum_{n\ge0} b_n (s-4)^n \right]
\eeq
Here $B$ is a variable and so are the $b_n$. Although the powers $(s-4)^n$ provide a basis in this interval, such functions grow as $s\rightarrow \infty$. To avoid this we consider the map we have already been using
\beq
z = \frac{\sqrt{4-s_a}-\sqrt{4-s}}{\sqrt{4-s_a}+\sqrt{4-s}}
\eeq
such that
\beq
\Im[(1-z)^{2k+1}] = (-1)^k\, 2^{2k+1} \left(\frac{s-4}{4-s_a}\right)^{k+\half} (1+\cO(s-4))
\eeq
Based on the previous parameterization we can write a parameterization that respects the leading behavior with functions that remain finite at $s\rightarrow\infty$ (or $z\rightarrow -1$):
\beqa
\rho_{0}^0 &=& -\frac{1}{(2\pi)^4} \frac{3}{32\pi} \left(1-\frac{s_a}{4}\right)^{\frac{1}{2}}  \Im\left\{(1-z)   \left[ A + (1-z)\sum_{n=0}^N a_n z^n\right] \right\}\\
\rho_{1}^1 &=& +\frac{1}{(2\pi)^4} \frac{1}{48\pi} \left(1-\frac{s_a}{4}\right)^{\frac{3}{2}} \Im\left\{   (1-z)^3 \left[ B + (1-z)\sum_{n=0}^N b_n z^n\right] \right\}  \\
\rho_{2}^0 &=& -\frac{1}{(2\pi)^4} \frac{1}{320\pi}\left(1-\frac{s_a}{4}\right)^{\frac{5}{2}} \Im\left\{  (1-z)^5 \left[ C + (1-z)\sum_{n=0}^N c_n z^n\right] \right\}
\eeqa 
The coefficients $A$,$B$,$C$ and $a_n$, $b_n$ and $c_n$ are real parameters. These parameterizations allow even powers of $(1-z)$ adding more flexibility.  Finally, to make sure that the leading term dominates, we regulate the coefficients by requiring $|a_n|\le a_{max}$, $|b_n|\le b_{max}$, $|c_n|\le c_{max}$. The value of the regulator should be chosen in a range that does not affect the results.

\section{Finite energy sum rules up to 3-loop}\label{fesr3}

The SVZ sum rules have played a major role in the understading of hadrons from QCD and therefore there is considerably material on them. In particular we quote here 3-loop results for the two-point correlator of the scalar and vector currents (keeping only terms with imaginary part). All expressions are for two flavors $N_f=2$ and taken from \cite{narison_2004} where one can find the original references:

\begin{equation}
\begin{aligned}
&\Pi^0_0(s) = -\frac{1}{(2\pi)^4}\frac{m_q^2}{8\pi^2} s \Bigg\{ 6\ln\left(-\frac{s}{\mu^2}\right) +\frac{\alpha_s}{\pi} \left[ 34\ln \left(-\frac{s}{\mu^2}\right)- 6 \ln^2\left(-\frac{s}{\mu^2}\right)\right]  \\
&  +\left(\frac{\alpha_s}{\pi}\right)^2 
\left[\left(\frac{10021}{24}-109\zeta(3)\right)\ln \left(-\frac{s}{\mu^2}\right)-\frac{296}{3} \ln^2 \left(-\frac{s}{\mu^2}\right)+\frac{53}{6}\ln^3 \left(-\frac{s}{\mu^2}\right)\right]+O(\alpha_s^3)\Bigg\}
\end{aligned}
\end{equation}
and
\begin{equation}
\begin{aligned}
\Pi^1_1 &=  -\frac{1}{(2\pi)^4}\frac{s}{8\pi^2} \Bigg\{\ln\left(-\frac{s}{\mu^2}\right) +\frac{\alpha_s}{\pi} \ln\left(-\frac{s}{\mu^2}\right)  \\
&  +\left(\frac{\alpha_s}{\pi}\right)^2 
\left[\left(\frac{107}{8}-\frac{29\zeta(3)}{3}\right)\ln \left(-\frac{s}{\mu^2}\right)-\frac{29}{24} \ln^2 \left(-\frac{s}{\mu^2}\right)\right]+O(\alpha_s^3) \Bigg\}
\end{aligned}
\end{equation}
In the previous expressions we only kept the parts that have a non-vanishing imaginary part because we are interested in $\rho = 2\,\Im \Pi(s+i\epsilon)$ to find the finite energy sum rules (setting $m_{\pi}=1$):
\begin{equation}
\begin{aligned}
&\frac{1}{s_0^{n+2}}\int_4^{s_0} \rho^0_0(x) x^n dx = \frac{m_q^2}{(2\pi)^4} \frac{3}{2\pi(n+2)} \Bigg\{1+\frac{\alpha_s}{\pi} \left[\frac{17}{3} + \frac{2}{n+2}\right] \\
  &+\left(\frac{\alpha_s}{\pi}\right)^2 
\bigg[\frac{10021}{144}-\frac{109}{6}\zeta(3)+\frac{296}{9(n+2)}+\frac{53}{6(n+2)^2}-\frac{53}{36}\pi^2\bigg]+O(\alpha_s^3)\Bigg\}
\end{aligned}
\end{equation}
and
\begin{equation}
\begin{aligned}
&\frac{1}{s_0^{n+2}}\int_4^{s_0} \rho^1_1(x) x^n dx=  \frac{1}{(2\pi)^4}\frac{1}{4\pi(n+2)} \Bigg\{1 +\frac{\alpha_s}{\pi}+\left(\frac{\alpha_s}{\pi}\right)^2 
\bigg[\left(\frac{107}{8}-\frac{29\zeta(3)}{3}\right)\\
& +\frac{29}{12(n+2)}\bigg]+O(\alpha_s^3) \Bigg\} 
\end{aligned}
\end{equation}
Finally, for the $S0$ and $P1$ waves, we find useful to improve the procedure by using a different analytical kernel such as the one in \cite{cherry2001qcd} that captures more on the lower energy region \footnote{as compared with the weight $s^n$ that tends to favor the larger values of $x$ close to $s_0$}:
\beq
K(s) = s^n \left(1-\frac{s}{s_0}\right) e^{-s/s_0}
\eeq
leading to the $S0,P1$ sum rules
\begin{subequations}\label{rhoc}
\begin{eqnarray}\label{rhoc1}
\frac{1}{s_0^{n+2}}\int_4^{s_0} \rho^0_0(x) K(x) dx &=& m_q^2 \left\{c^{S0}_{n,0} + c^{S0}_{n,1} \frac{\alpha_s}{\pi} + c^{S0}_{n,2} \left(\frac{\alpha_s}{\pi}\right)^2+\cO(\alpha_s^3) \right\}\\ \label{rhoc2}
\frac{1}{s_0^{n+2}}\int_4^{s_0} \rho^1_1(x) K(x) dx &=& c^{P1}_{n,0} + c^{P1}_{n,1} \frac{\alpha_s}{\pi} + c^{P1}_{n,2} \left(\frac{\alpha_s}{\pi}\right)^2+\cO(\alpha_s^3) 
\end{eqnarray}
\end{subequations}
In the main text we use these ``pinched" sum rules for $\rho^0_0,\rho^1_1$ up to $n=6$. (For $\rho^0_2$ we use the same sum rules with weights $s^n$ as in previous work \cite{He:2024nwd}.) The corresponding numerical coefficients $c_{i,j}$ are shown in tables \ref{tab:coefficients1}, \ref{tab:coefficients2} for reference. The numerical values are used to constrain the spectral densities following eq.(\ref{rhoc}).
\begin{table}[h]
    \centering
    \begin{tabular}{c|ccc}
        \( i \backslash j \) & 0 & 1 & 2 \\
        \hline
        0 & \( 3.1750 \times 10^{-5} \) & \( 2.4104 \times 10^{-4} \) & \( 2.2490 \times 10^{-3} \) \\
        1 & \( 1.4299 \times 10^{-5} \) & \( 1.0012 \times 10^{-4} \) & \( 8.3028 \times 10^{-4} \) \\
        2 & \( 7.9936 \times 10^{-6} \) & \( 5.3436 \times 10^{-5} \) & \( 4.1288 \times 10^{-4} \) \\
        3 & \( 5.0653 \times 10^{-6} \) & \( 3.2859 \times 10^{-5} \) & \( 2.4218 \times 10^{-4} \) \\
        4 & \( 3.4828 \times 10^{-6} \) & \( 2.2124 \times 10^{-5} \) & \( 1.5766 \times 10^{-4} \) \\
        5 & \( 2.5359 \times 10^{-6} \) & \( 1.5862 \times 10^{-5} \) & \( 1.1022 \times 10^{-4} \) \\
    \end{tabular}
    \caption{Numerical values of the S0 sum rule coefficients \( c^{S0}_{i,j} \). These coefficients are used to constrain the spectral density}
    \label{tab:coefficients1}
\end{table}

\begin{table}[h]
    \centering
    \begin{tabular}{c|ccc}
        \( i \backslash j \) & 0 & 1 & 2 \\
        \hline
        -1 & \( 1.8783 \times 10^{-5} \) & \( 1.8783 \times 10^{-5} \) & \( 1.1097 \times 10^{-4} \) \\
         0 & \( 5.2916 \times 10^{-6} \) & \( 5.2916 \times 10^{-6} \) & \( 2.1597 \times 10^{-5} \) \\
         1 & \( 2.3831 \times 10^{-6} \) & \( 2.3831 \times 10^{-6} \) & \( 8.0286 \times 10^{-6} \) \\
         2 & \( 1.3323 \times 10^{-6} \) & \( 1.3323 \times 10^{-6} \) & \( 3.9775 \times 10^{-6} \) \\
         3 & \( 8.4421 \times 10^{-7} \) & \( 8.4421 \times 10^{-7} \) & \( 2.3186 \times 10^{-6} \) \\
         4 & \( 5.8047 \times 10^{-7} \) & \( 5.8047 \times 10^{-7} \) & \( 1.4997 \times 10^{-6} \) \\
    \end{tabular}
    \caption{Numerical values of the P1 sum rule coefficients \( c^{P1}_{i,j} \).}
    \label{tab:coefficients2}
\end{table}

\bigskip

\bigskip

\bigskip

\bibliographystyle{utphys}
\bibliography{references}

\end{document}